\documentclass[twocolumn]{aastex62}
\usepackage{appendix}
\usepackage{natbib}
\usepackage{amsmath}
\usepackage{booktabs}
\usepackage{color,subfigure,lineno} 
\usepackage{ulem}

\usepackage{lineno}
\newcommand{\SiII}{\makebox{[Si{\sc\,II}]\,}}

\newcommand{\lya}{Ly\,$\alpha$}
\newcommand{\CIV}{C\,{\sevenrm IV}}

\newcommand{\SiIV}{Si\,{\sevenrm IV}}
\newcommand{\CIII}{C\,{\sevenrm III]}}

\newcommand{\AlIII}{Al\,{\sevenrm III}}

\newcommand{\SiIII}{Si\,{\sevenrm III]}}

\def\FeII{Fe\,{\sc ii}}
\def\FeIII{Fe\,{\sc iii}}
\def\MgII{Mg\,{\sc ii}}
\def\HeII{He\,{\sc ii}}

\newcommand{\OIIIUV}{O{\sevenrm\,III}]}
\def \OIII {[O\,{\sc iii}]}

\newcommand{\NIII}{N\,{\sevenrm\,III}]}

\def \NV {N\,{\sc v}}
\newcommand{\SII}{[S{\sevenrm\,II}]}

\newcommand{\OIV}{O\,{\sevenrm IV]}}
\newcommand{\OVI}{O\,{\sevenrm VI}}

   \font\sevenrm=cmr7 scaled 1000
\newcommand{\comments}[1]{}

\begin{document}
\title{Metallicity in Quasar Broad Line Regions at Redshift $\sim6$}
%\author{Shu Wang, Linhua Jiang,  Yue Shen et al. }

\author[0000-0002-2052-6400]{Shu Wang}
\affiliation{Kavli Institute for Astronomy and Astrophysics, Peking University, Beijing 100871, China;  wangshukiaa@pku.edu.cn, jiangKIAA@pku.edu.cn}
%\affiliation{Department of Astronomy, School of Physics, Peking University, Beijing 100871, China}
\affiliation{Astronomy Program, Department of Physics and Astronomy, Seoul National University, Seoul, 08826, Republic of Korea}
%\email{wangshukiaa@pku.edu.cn, jiangKIAA@pku.edu.cn}

\author[0000-0003-4176-6486]{Linhua Jiang}
%\affiliation{Kavli Institute for Astronomy and Astrophysics, Peking University, Beijing 100871, China}
\affiliation{Kavli Institute for Astronomy and Astrophysics, Peking University, Beijing 100871, China;  wangshukiaa@pku.edu.cn, jiangKIAA@pku.edu.cn}
\affiliation{Department of Astronomy, School of Physics, Peking University, Beijing 100871, China}
%\email{jiangKIAA@pku.edu.cn}

\author[0000-0002-6893-3742]{Yue Shen}
\affiliation{Department of Astronomy, University of Illinois at Urbana-Champaign, Urbana, IL 61801, USA}
\affiliation{National Center for Supercomputing Applications, University of Illinois at Urbana-Champaign, Urbana, IL 61801, USA}

\author[0000-0001-6947-5846]{Luis C. Ho}
%\affiliation{Kavli Institute for Astronomy and Astrophysics, Peking University, Beijing 100871, China}
\affiliation{Kavli Institute for Astronomy and Astrophysics, Peking University, Beijing 100871, China;  wangshukiaa@pku.edu.cn, jiangKIAA@pku.edu.cn}
\affiliation{Department of Astronomy, School of Physics, Peking University, Beijing 100871, China}

\author[0000-0001-9191-9837]{Marianne Vestergaard}
\affiliation{Steward Observatory, University of Arizona, 933 North Cherry Avenue, Tucson, AZ 85721-0065, USA}
\affiliation{The Niels Bohr Institute at University of Copenhagen, Juliane Maries Vej 30, DK-2100 Copenhagen, Denmark}

\author[0000-0002-2931-7824]{Eduardo Ba{\~n}ados} 
\affiliation{Max Planck Institut f\"ur Astronomie, K\"onigstuhl 17, D-69117, Heidelberg, Germany}

\author[0000-0002-4201-7367]{Chris J. Willott}
\affiliation{NRC Herzberg, 5071 West Saanich Rd, Victoria, BC V9E 2E7, Canada}

\author[0000-0002-6168-3867]{Jin Wu}
%\affiliation{Kavli Institute for Astronomy and Astrophysics, Peking University, Beijing 100871, China}
\affiliation{Kavli Institute for Astronomy and Astrophysics, Peking University, Beijing 100871, China;  wangshukiaa@pku.edu.cn, jiangKIAA@pku.edu.cn}

\author[0000-0002-3983-6484]{Siwei Zou}
%\affiliation{Kavli Institute for Astronomy and Astrophysics, Peking University, Beijing 100871, China}
\affiliation{Kavli Institute for Astronomy and Astrophysics, Peking University, Beijing 100871, China;  wangshukiaa@pku.edu.cn, jiangKIAA@pku.edu.cn}

\author[0000-0001-5287-4242]{Jinyi Yang}
\altaffiliation{Strittmatter Fellow} 
\affiliation{Steward Observatory, University of Arizona, 933 North Cherry Avenue, Tucson, AZ 85721-0065, USA}

\author[0000-0002-7633-431X]{Feige Wang}
\altaffiliation{NHFP Hubble Fellow}
\affiliation{Steward Observatory, University of Arizona, 933 North Cherry Avenue, Tucson, AZ 85721-0065, USA}

\author[0000-0003-3310-0131]{Xiaohui Fan}
\affiliation{Steward Observatory, University of Arizona, 933 North Cherry Avenue, Tucson, AZ 85721-0065, USA}

\author[0000-0002-7350-6913]{Xue-Bing Wu}
%\affiliation{Kavli Institute for Astronomy and Astrophysics, Peking University, Beijing 100871, China}
\affiliation{Kavli Institute for Astronomy and Astrophysics, Peking University, Beijing 100871, China;  wangshukiaa@pku.edu.cn, jiangKIAA@pku.edu.cn}
\affiliation{Department of Astronomy, School of Physics, Peking University, Beijing 100871, China}

%\author[0000-0001-9585-1462]{Dominik A. Riechers}
%\affiliation{Cornell University, Space Sciences Building, Ithaca, NY 14853, USA}

%\author[0000-0001-9024-8322]{Bram Venemans}
%\affiliation{Max Planck Institut f\"{u}r Astronomie, K\"{o}nigstuhl 17, D-69117, Heidelberg, Germany}

%\author[0000-0003-4793-7880]{Fabian Walter}
%\affiliation{Max Planck Institut f\"{u}r Astronomie, K\"{o}nigstuhl 17, D-69117, Heidelberg, Germany}

%\author[0000-0002-5535-4186]{Ravi Joshi}
%\affiliation{Kavli Institute for Astronomy and Astrophysics, Peking University, Beijing 100871, China}

\begin{abstract}
Broad line regions (BLRs) in high-redshift quasars provide crucial information of chemical enrichment in the early universe. Here we present a study of BLR metallicities in 33 quasars at redshift $5.7<z<6.4$. Using the near-IR spectra of the quasars obtained from the Gemini telescope, we measure their rest-frame UV emission line flux and calculate flux ratios. We then estimate BLR metallicities with empirical calibrations based on photoionization models. The inferred median metallicity of our sample is a few times the solar value, indicating that the BLR gas had been highly metal-enriched at $z\sim6$. We compare our sample with a low-redshift quasar sample with similar luminosities and find no evidence of redshift evolution in quasar BLR metallicities. This is consistent with previous studies. The \FeII$/$\MgII\ flux ratio, a proxy for the Fe$/\alpha$ element abundance ratio, shows no redshift evolution as well, further supporting rapid nuclear star formation at $z\sim6$. We also find that the black hole mass-BLR metallicity relation at $z\sim6$ is consistent with the relation measured at $2<z<5$, suggesting that our results are not biased by a selection effect due to this relation.   
\end{abstract}


\section{Introduction}    
The evolution of metallicity across cosmic time contains crucial information about the star formation history and galaxy evolution. Studies of metallicity in different environments, including star-forming galaxies ($0<z<3.5$) \citep[e.g.,][]{Tremonti04,Maiolino08,Mannucci09,Zahid14a,Bian17}, Damped Lyman-$\alpha$ systems (DLAs) ($0<z<5$) \citep[e.g.,][]{Ledoux06,Moller13, Prochaska13,Rafelski14, Banados19}, \lya\ emitters ($0<z<3$)  \citep[e.g.,][]{Nakajima12,Nakajima13,Guo20b}, and quasars (up to $z\sim7$) \citep[e.g.,][]{Dietrich03,Nagao06,Jiang07,DeRosa11,Xu18,Onoue20}, have been carried out in the past two decades.

To measure metallicity in the early universe ($z\gtrsim 5.7$) and explore the reionization epoch, luminous Active Galactic Nuclei (AGNs), i.e., quasars, are among the most valuable probes given their high luminosity. In recent years, significant progress has been made in discovering high-redshift quasars \citep[e.g.,][]{Fan06,Willott07,Willott10a,Mortlock11,Jiang16,Banados16,Banados18,Matsouka18b,Matsouka19,Reed19,Yang19,Yang20,Wang21}. Near-infrared (NIR) spectroscopy of these earliest quasars reveals that the black holes (BHs) are already massive ($10^{9}\sim 10^{10}{\rm M}_{\odot}$) \citep{Jiang07,Kurk07,Wu15,Shen19a}. Studies of the BLR metallicities in these high-redshift quasars found no evidence of strong evolution compared with their low-redshift counterparts \citep{Hamann93,Barth03,Dietrich03c,Dietrich03,Dietrich03b,Nagao06,Kurk07,Jiang07,Juarez09,DeRosa11,DeRosa14,Mazzucchelli17,Xu18,Shen19a,Shin19,Tang19,Schindler20,Onoue20}. 

Measurement of quasar BLR metallicities mainly relies on the rest-frame UV broad emission lines. Photoionization calculations suggest that UV line flux ratios, including (\SiIV+\OIV)$/$\CIV, (\CIII+\SiIII)$/$\CIV, \AlIII$/$\CIV, \NV$/$\CIV, \HeII$/$\CIV, and \NV$/$\HeII\ can be used to infer metallicities \citep{Hamann02, Nagao06}. \citet{Nagao06} used a large quasar sample from the SDSS, and measured different diagnostic flux ratios from the composite spectra in each redshift and absolute B magnitude ($M_B$) bin. No significant redshift evolution was found in the redshift range of $2<z<4.5$. By comparing the observed line flux ratios with their photoionization calculations, the BLR metallicities are estimated to be ${\rm Z}\sim 5{\rm Z}_{\odot}$ (solar metallicity). \citet{Jiang07} performed a similar analysis for six quasars at $z\sim6$, and the BLR metallicity is found to be consistent with those in low-redshift quasars (${\rm Z}\sim 4{\rm Z}_{\odot}$).  

On the other hand, a significant positive correlation between BLR metallicity and quasar luminosity is found in \citet{Nagao06}. As suggested by the paper, the more fundamental relation is the mass-metallicity relation (MZR) in AGNs \citep{Matsouka11,Xu18}, i.e., more massive BHs are accompanied by BLRs with higher metallicities. The origin of the MZR relation is unclear, but is likely connected to the galaxy MZR via the BH mass-host galaxy relation \citep{Matsouka11,Matsouka18,Dors15}. 

Another critical finding in quasar metallicity studies is the non-redshift evolution of \FeII$/$\MgII,  which is the first-order proxy of the Fe$/\alpha$ element abundance ratio. This ratio acts like a clock of the star formation history, since the Fe element is mainly ejected by Type Ia supernovae (SN Ia), which is delayed by $\sim$1 Gyr relative to $\alpha$ elements that are mainly produced by core collapsed supernovae (Type II and Type Ib/Ic supernovae). Many studies measured the \FeII$/$\MgII\ ratio \citep{Dietrich03,Kurk07,Jiang07,DeRosa11, DeRosa14,Mazzucchelli17,Shin19,Schindler20,Onoue20, Yang21}, and no redshift evolution was found up to $z\sim7$. This result suggests a rapid star formation in the nucleus at earlier epochs to produce the observed iron abundance in $z\sim 6$ quasar BLRs.

\citet{Shen19a} conducted a large spectroscopic survey of 50 $z\ge 5.7$ quasars using Gemini GNIRS with a simultaneous coverage in NIR. This sample is a valuable data set to study the physical properties of high-redshift quasars as well as intervening absorption systems that trace the intergalactic medium (IGM) and the circum-galactic medium (CGM) \citep{Zou20}. \citet{Shen19a} performed an initial analysis of the sample, focusing on BH masses, emission line shifts, and other general spectral properties. They found that the median composite spectrum of $z\sim 6$ quasars is similar to that generated from a luminosity-matched control sample at lower redshifts.

In this paper, we perform a quantitative analysis to measure quasar BLR metallicity of the GNIRS sample and study its redshift evolution. Although similar analyses have been performed in the past, our GNIRS sample has the advantages of better sample statistics, and a more complete coverage of multiple UV broad lines to measure the BLR metallicity with different indicators. The paper is organized as follows. We describe the observations and sample selection in \S2. The detailed spectral decomposition is presented in \S3. In \S4 we study the redshift evolution of BLR metallicities and the \FeII$/$\MgII\ ratio. We discuss the implications of our results in \S5 and summarize in \S6. Throughout this paper, we adopt a flat lambda cold dark matter cosmology with $\Omega_0 = 1 - \Omega_{\Lambda} = 0.3$ and ${\rm H}_0=70\;{\rm km}/{\rm s}\;{\rm Mpc}^{-1}$.

\section{Sample and Data}

We use the high-redshift quasars from \citet{Shen19a} as our parent sample. This parent sample is collected from identified quasars at $z\geq5.7$ in the literature. 51 quasars were observed during the 15B-17A semesters using GNIRS on Gemini-North. One object was removed from the sample because of poor observing conditions, and the final parent sample contains 50 quasars. The properties of the parent sample, including coordinates, $J$ band photometry, and the discovery references, are summarized in Table 1 of \citet{Shen19a}. The parent sample is not a complete flux-limited sample at $z\geq5.7$, but it includes quasars with diverse properties in terms of luminosity and spectral properties.

The GNIRS observations were conducted in the cross-dispersion mode using the short blue camera with a slit width of $0$\farcs$675$. The wavelength coverage of the spectroscopy is 0.85-2.5$~\mu$m, with a spectral resolution of $R\sim650$. The exposure time varies from 30 minutes to 5 hours depending on the brightness of the target. The obtained signal-to-noise ratio (S/N) is 5 per pixel averaged over $H$ band.

The GNIRS data were reduced using a custom pipeline based on two existing pipelines for GNIRS: the PyRAF-based XDGNIRS \citep{Mason15} and the IDL-based XIDL package\footnote{{http://www.ucolick.org/$\sim$xavier/IDL/}}. The spectrum is scaled to the available $J$ band magnitude for absolute flux calibration. Detailed data reduction is described in Section 2.1 of \citet{Shen19a}. 

Starting from this parent sample, we perform an initial spectral analysis as described in \S\ref{sec:spectral_analysis}. We found that some quasars have low S/N, peculiar continuum shapes likely caused by reduction or intrinsic reddening, or significantly affected by strong telluric line residuals. To robustly measure \CIV\ based metallicity and \FeII$/$\MgII, we exclude objects for which neither \CIV\ nor \MgII\ can be reasonably fitted (see \S\ref{sec:spectral_analysis}). The final sample for our study contains 33 quasars.

\section{Spectral analysis}\label{sec:spectral_analysis}

\begin{figure*}[t]
   \centering
   \includegraphics[width=1.0\textwidth]{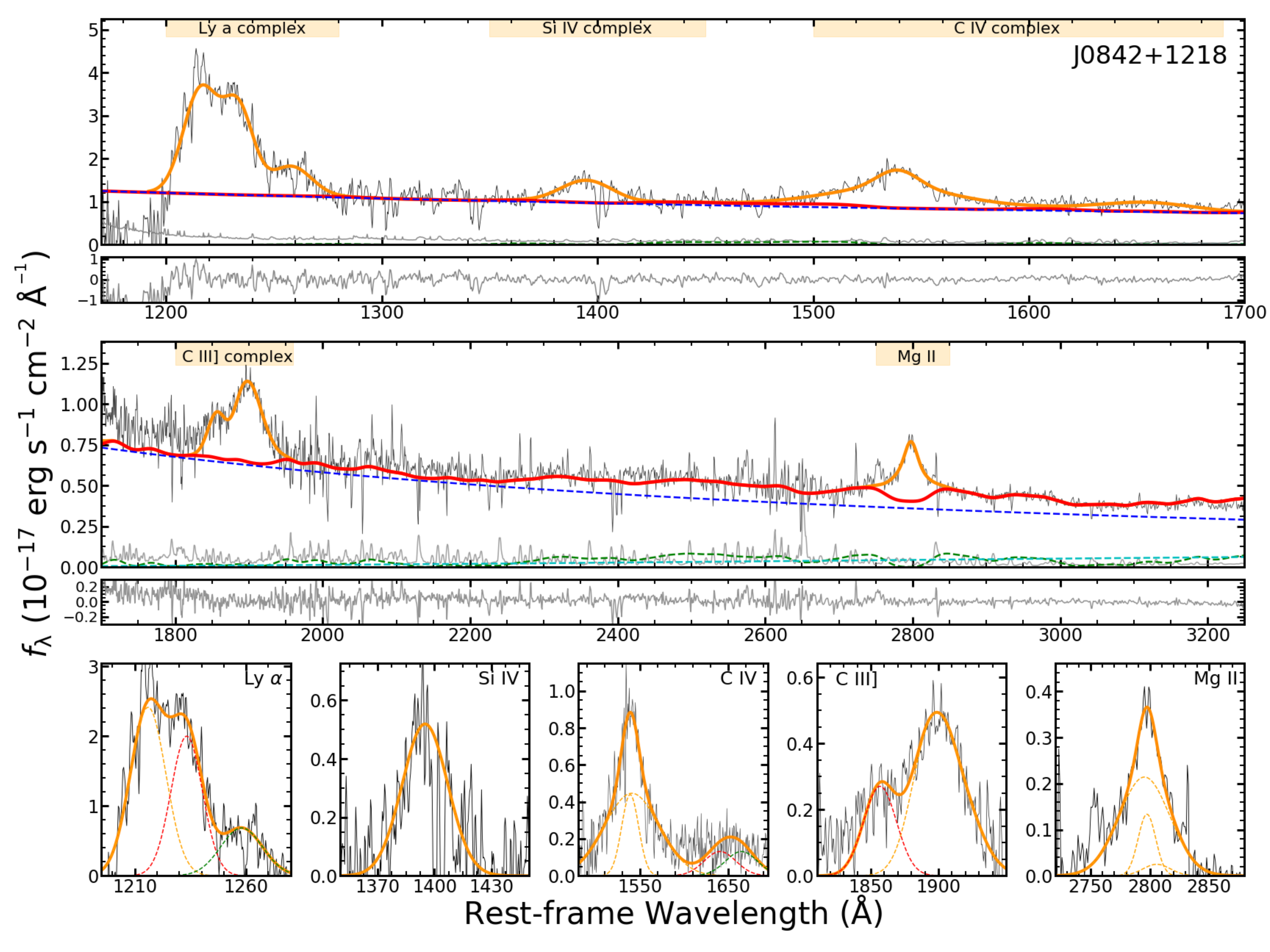} 
    \caption{An example of our spectral analysis. The upper panels present the global continuum fitting results and the residuals in two spectral regions. The black and grey lines represent the original flux and error spectrum, respectively. The red and orange solid lines are the pseudo-continuum and emission line models, respectively. Different line complexes are labeled on the top and the light orange shaded areas denote the wavelength ranges to fit each complex. The bottom-row panels display the fits to the pseudo-continuum-subtracted line profiles in different line complexes, \lya, \SiIV, \CIV, \CIII, and \MgII, from left to right. The orange, red and green dashed Gaussians represent the main components in each complex, decomposed components used to indicate metallicity, and those not used in metallicity estimation but included in the fitting, respectively. The complete figure set for 33 objects are available in the online journal. }
 
%\figsetstart
%\figsetnum{1}
%\figsettitle{Complete figure set for the fits of all the 33 objects.}
\figsetgrpnum{1.1}
\figsetgrptitle{J0002$+$2550}
\figsetplot{Figures1_figureset/fitting_J0002$+$2550.pdf}
\figsetgrpend
\figsetgrpnum{1.2}
\figsetgrptitle{J0008$-$0626}
\figsetplot{Figures1_figureset/fitting_J0008$-$0626.pdf}
\figsetgrpend
\figsetgrpnum{1.3}
\figsetgrptitle{J0028$+$0457}
\figsetplot{Figures1_figureset/fitting_J0028$+$0457.pdf}
\figsetgrpend
\figsetgrpnum{1.4}
\figsetgrptitle{J0050$+$3445}
\figsetplot{Figures1_figureset/fitting_J0050$+$3445.pdf}
\figsetgrpend
\figsetgrpnum{1.5}
\figsetgrptitle{J0353$+$0104}
\figsetplot{Figures1_figureset/fitting_J0353$+$0104.pdf}
\figsetgrpend
\figsetgrpnum{1.6}
\figsetgrptitle{J0810$+$5105}
\figsetplot{Figures1_figureset/fitting_J0810$+$5105.pdf}
\figsetgrpend
\figsetgrpnum{1.7}
\figsetgrptitle{J0835$+$3217}
\figsetplot{Figures1_figureset/fitting_J0835$+$3217.pdf}
\figsetgrpend
\figsetgrpnum{1.8}
\figsetgrptitle{J0836$+$0054}
\figsetplot{Figures1_figureset/fitting_J0836$+$0054.pdf}
\figsetgrpend
\figsetgrpnum{1.9}
\figsetgrptitle{J0840$+$5624}
\figsetplot{Figures1_figureset/fitting_J0840$+$5624.pdf}
\figsetgrpend
\figsetgrpnum{1.10}
\figsetgrptitle{J0841$+$2905}
\figsetplot{Figures1_figureset/fitting_J0841$+$2905.pdf}
\figsetgrpend
\figsetgrpnum{1.11}
\figsetgrptitle{J1044$-$0125}
\figsetplot{Figures1_figureset/fitting_J1044$-$0125.pdf}
\figsetgrpend
\figsetgrpnum{1.12}
\figsetgrptitle{J1137$+$3549}
\figsetplot{Figures1_figureset/fitting_J1137$+$3549.pdf}
\figsetgrpend
\figsetgrpnum{1.13}
\figsetgrptitle{J1143$+$3808}
\figsetplot{Figures1_figureset/fitting_J1143$+$3808.pdf}
\figsetgrpend
\figsetgrpnum{1.14}
\figsetgrptitle{J1148$+$0702}
\figsetplot{Figures1_figureset/fitting_J1148$+$0702.pdf}
\figsetgrpend
\figsetgrpnum{1.15}
\figsetgrptitle{J1148$+$5251}
\figsetplot{Figures1_figureset/fitting_J1148$+$5251.pdf}
\figsetgrpend
\figsetgrpnum{1.16}
\figsetgrptitle{J1207$+$0630}
\figsetplot{Figures1_figureset/fitting_J1207$+$0630.pdf}
\figsetgrpend
\figsetgrpnum{1.17}
\figsetgrptitle{J1243$+$2529}
\figsetplot{Figures1_figureset/fitting_J1243$+$2529.pdf}
\figsetgrpend
\figsetgrpnum{1.18}
\figsetgrptitle{J1250$+$3130}
\figsetplot{Figures1_figureset/fitting_J1250$+$3130.pdf}
\figsetgrpend
\figsetgrpnum{1.19}
\figsetgrptitle{J1257$+$6349}
\figsetplot{Figures1_figureset/fitting_J1257$+$6349.pdf}
\figsetgrpend
\figsetgrpnum{1.20}
\figsetgrptitle{J1429$+$5447}
\figsetplot{Figures1_figureset/fitting_J1429$+$5447.pdf}
\figsetgrpend
\figsetgrpnum{1.21}
\figsetgrptitle{J1436$+$5007}
\figsetplot{Figures1_figureset/fitting_J1436$+$5007.pdf}
\figsetgrpend
\figsetgrpnum{1.22}
\figsetgrptitle{J1545$+$6028}
\figsetplot{Figures1_figureset/fitting_J1545$+$6028.pdf}
\figsetgrpend
\figsetgrpnum{1.23}
\figsetgrptitle{J1602$+$4228}
\figsetplot{Figures1_figureset/fitting_J1602$+$4228.pdf}
\figsetgrpend
\figsetgrpnum{1.24}
\figsetgrptitle{J1609$+$3041}
\figsetplot{Figures1_figureset/fitting_J1609$+$3041.pdf}
\figsetgrpend
\figsetgrpnum{1.25}
\figsetgrptitle{J1623$+$3112}
\figsetplot{Figures1_figureset/fitting_J1623$+$3112.pdf}
\figsetgrpend
\figsetgrpnum{1.26}
\figsetgrptitle{J1630$+$4012}
\figsetplot{Figures1_figureset/fitting_J1630$+$4012.pdf}
\figsetgrpend
\figsetgrpnum{1.27}
\figsetgrptitle{J2310$+$1855}
\figsetplot{Figures1_figureset/fitting_J2310$+$1855.pdf}
\figsetgrpend
\figsetgrpnum{1.28}
\figsetgrptitle{P000+$2$6}
\figsetplot{Figures1_figureset/fitting_P000+$2$6.pdf}
\figsetgrpend
\figsetgrpnum{1.29}
\figsetgrptitle{P060+$2$4}
\figsetplot{Figures1_figureset/fitting_P060+$2$4.pdf}
\figsetgrpend
\figsetgrpnum{1.30}
\figsetgrptitle{P210+$2$7}
\figsetplot{Figures1_figureset/fitting_P210+$2$7.pdf}
\figsetgrpend
\figsetgrpnum{1.31}
\figsetgrptitle{P228+$2$1}
\figsetplot{Figures1_figureset/fitting_P228+$2$1.pdf}
\figsetgrpend
\figsetgrpnum{1.32}
\figsetgrptitle{P333+$2$6}
\figsetplot{Figures1_figureset/fitting_P333+$2$6.pdf}
\figsetgrpend
%\figsetend

   \label{fig:fitting}
\end{figure*}

We fit the NIR spectra with multi-component models, following earlier work \citep[e.g.,][]{Shen11, Shen19b, Wang20}. In the fitting procedure, we correct the effects of narrow absorption lines with iterative sigma clipping. We visually check the initial fitting results, and interactively add additional pixel masks to exclude broad absorption features in the fits on an object-by-object basis. We exclude individual line complex regions if, e.g., the underlying continuum has peculiar shapes and/or the absorption features are too broad to be fully masked. In addition, we mask the spectral regions where the spectrum is heavily affected by telluric absorption. These telluric regions are generally $\lambda_{\rm obs}=13400-14500$\,\AA\ and $\lambda_{\rm obs}=17800-19000$\,\AA, but are slightly adjusted based on the initial fitting results.

After the initial fits, we perform the final fits in the rest-frame of the quasar using the systemic redshifts, $z_{\rm sys}$, provided in \citet{Shen19a}. An example of our spectral modeling (J0842+1218) is shown in Figure \ref{fig:fitting}. The complete figure set of our fits is available in the online journal.

During the spectral modeling, we first fit a pseudo-continuum model to several emission-line-free windows, and then fit each line complex after subtracting the pseudo-continuum model. These line complexes are \lya$+$\NV, \SiIV$+$\OIV, \CIV$+$\HeII$+$\OIII, \CIII$+$\SiIII$+$\AlIII, and \MgII\ complex. The pseudo-continuum model is described in \S\ref{sec:continuum_model}, and the emission-line models are described in \S\ref{sec:emissionline_model}. \S\ref{sec:spectral_measurements} summarizes our spectral measurements and derived physical quantities, including BH masses, bolometric luminosities, and Eddington ratios.

\subsection{The continuum model} \label{sec:continuum_model}

We fit a global continuum in order to measure different line flux ratios consistently across the spectrum. We choose the commonly used continuum windows in the rest-frame UV, which consist of several emission-line free wavelength regions, including $\lambda_{\rm rest}$ = $1345-1350 $, $1445-1450$, $1700-1705$. $2150-2400$, $2480-2675$, and $2900-3450$\,\AA. The continuum models used in our fits consist of a power law component $F_{\rm PL}$, a Balmer continuum component $F_{\rm BC}$, and a broad \FeII\ emission component $F_{\rm Fe}$. Different from \citet{Shen19a}, we do not add an additional polynomial component to account for the peculiar continuum shapes in some objects caused by intrinsic reddening or flux calibration issues. This component would introduce additional systematic uncertainties to the \FeII$/$\MgII\ measurement. Instead, we exclude the object from our sample if we cannot fit the original spectrum well by only including the model components described above, as judged by visual inspection.

The power law model $F_{\rm PL}$ is normalized at 3000\,\AA, and described by:
\begin{equation}
    F_{\rm PL}(\lambda ;\;\alpha,\beta) = \alpha \left(\frac{\lambda}{3000}\right)^{\beta}\ ,
\end{equation}
where $\alpha$ and $\beta$ are the flux scaling factor and the power-law slope, respectively.

The Balmer continuum model $F_{\rm BC}$ is described by the following equation:
\begin{equation}
    F_{\rm BC} = \gamma \, B_{\lambda}(\lambda, T_e)\;(1 - e^{-\tau_{\lambda}}) \;; \lambda \leq \lambda_{\rm BL}, 
\end{equation}
where $\gamma$ is the flux scaling factor and $B_{\lambda}(\lambda, T_e)$ is the Plank function at temperature $T_e$. $\lambda_{\rm BL}=3646$\,\AA\ is the wavelength of Balmer limit. $\tau_{\lambda} = \tau_{\rm BL} (\lambda/\lambda_{\rm BL})^3$ is the optical depth normalized by the value of Balmer limit. The normalization of the Balmer continuum is commonly determined at $\lambda=3675$\,\AA, where there is no obvious contamination from iron emission. However, because of the limited spectral coverage of our spectrum, $F_{\rm BC}$ and $F_{\rm PL}$ are highly degenerated in our fits. Thus, we follow previous studies \citep{Dietrich03,Kurk07,DeRosa11,Mazzucchelli17,Shin19,Schindler20,Onoue20} and set the flux density of Balmer continuum component to be 30\% of that of power law component at the Balmer limit, as described by the following equation:
\begin{equation}
    F_{\rm BC}(\lambda=3646) = 0.3 \times F_{\rm PL}(\lambda=3646).
\end{equation}

The \FeII\ pseudo-continuum $F_{\rm Fe}$ is fitted with an empirical template $F_{\rm Template}$. The free parameters include the flux scaling factor, the width of the broadening kernel, and the wavelength shift, as described by:
\begin{equation}
    F_{\rm Fe} = \zeta F_{\rm Template}|_{\lambda \times (1+\delta)} \circledast G(\lambda, \sigma),  \label{eq:4}
\end{equation}
where $\zeta$ is the flux scaling factor, $G(\lambda, \sigma)$ is a Gaussian broadening kernel with a kernel width $\sigma$, and $\delta$ is the wavelength shift parameter.

One of the commonly used \FeII\ templates in the rest-frame UV wavelength range was provided by \citet{VestergaardWilkes01}, hereafter the VW01 template. The VW01 template was constructed based on the high S/N HST spectra of a narrow-line Seyfert 1 AGN, I Zwicky 1. In the wavelength region beneath \MgII\ ($2780-2830$\,\AA), the flux of the VW01 template is set to be zero. \citet{Tsuzuki06} presented a \FeII\ template (hereafter the T06 template) that uses information from photoionization calculations \citep{Ferland98}. The two templates are generally consistent with each other. However, \citet{Woo18} compared the fitting results based on these two templates and found that the \MgII\ (\FeII) flux using the VW01 template is systematically higher (lower) compared to the measurements based on the T06 template. This will introduce a systematic bias when we compare \FeII$/$\MgII\ ratios from the literature using different \FeII\ templates.

In this work, we adopt the VW01 template, since many empirical scaling relations and BH mass estimators are based on this template. We compare with the fitting results using the T06 template in the \ref{sec:appendix_1}. Note that the T06 template only covers the wavelength region of $\lambda \geq 2200$\,\AA. In the $\lambda<2200$\,\AA\ region, the T06 template is augmented by the VW01 template. Therefore, the choice of the \FeII\ template will only affect the \FeII$/$\MgII\ ratio, but have little effect on other UV line flux ratios. This is also supported by \citet{Onoue20} (see their Table 3).

\subsection{Emission line models} \label{sec:emissionline_model}

We focus on main metallicity diagnostic lines including \CIV, \HeII, \SiIV, \NV, \CIII, \AlIII, and \MgII. Many of them are  blended with other emission lines, e.g., \NV\ $\lambda1240$ with \lya, \SII\  $\lambda1256$, and \SiII\ $\lambda1263$, \SiIV\ $\lambda\lambda 1394, 1403$ with \OIV\ $\lambda1402$, and \CIII\ $\lambda\lambda 1907, 1909$ with \SiIII\ $\lambda1872$. Therefore, we attempt to decompose these lines in individual line complexes. Table \ref{tab:fitting_parameter} summarizes all lines included in our spectral analysis and the fitting ranges for each line complex.

We use multiple Gaussians to fit the line profile in each complex. Some earlier works also used a broken power-law model \citep[e.g.,][]{Nagao06,Matsouka11,Xu18} to describe the line profile. They divide lines into high and low ionization line groups and assume all lines in the same group have the same power-law indices and line shifts. This can be over-simplified, e.g., \CIII\ and \MgII\ may have different line shifts \citep{Shen16}. \citet{Nagao06} also found that fitting results using multiple Gaussians are as good as those from the broken power-law model. Table \ref{tab:fitting_parameter} shows the number of Gaussians used in our fits. We generally do not include UV narrow line components given the spectral quality, unless they are relatively strong and obvious, e.g., \MgII\ lines for some quasars. The inclusion of narrow lines in such cases is important, since it affects the measurement of the broad-line  full-width-at-half-maximum (FWHM), which is used in the BH mass estimation.

\begin{table}[t] 
   \caption{Fitting parameters}     \label{tab:fitting_parameter}
    \tabcolsep 4pt
    \centering
    \small

    \begin{tabular}{ c c c c}
     
    \hline \hline
    Line complex &  Fitting range &  Line & ${{\rm n}_{\rm gauss}}^a$ \\ \hline
    \MgII\       &   $2750-2850$ & \MgII\ & 3B$+$2N \\ 
    \CIII\       &  $1800-1960$ & \CIII$+$\SiIII\  & 2B \\ 
                 &   & \AlIII\ & 1B \\
    \CIV\        &  $1500-1690$ & \CIV\ & 3B \\ 
                 &   & \HeII\  & 1B \\
                 &   & \OIV\   & 1B \\
    \SiIV\       &  $1350-1450$  & \SiIV$+$\OIV\  & 2B \\
     \lya\       &  $1200-1280$ & \lya\ & 2B+1N \\
                 &   & \NV\    & 1B \\
                 &   & \SII$+$\SiII\   & 1B \\
    \hline
    
    \multicolumn{4}{p{0.45\textwidth}}{Note. a. In the last column, B and N refer to the Gaussians used for the broad- and narrow-line components, respectively.}
\end{tabular}
\end{table}

Apart from being blended with \FeII, the main difficulty in fitting \MgII\ is the contamination from telluric absorptions. For the typical redshift of our sample ($z\sim6$), \MgII\ shifts to 19,600\,\AA\ in the observed frame. The blue wing of \MgII, or even most of the line profile, can be affected by the telluric absorption. Therefore, we adjust the blue side boundary when necessary for individual objects. We exclude some objects whose \MgII\ lines are severely affected by the telluric absorption.

\CIII\ can also be affected by telluric absorption. However, we find that the telluric absorption within $13,400<\lambda_{\rm obs}<14,500$\,\AA\ usually only affects a very small portion of the \CIII\ profile, and many of the affected objects are already excluded from the continuum fitting step.  Only one \CIII\ fit suffered from severe telluric effects and was excluded from our analysis. In the fitting of \CIII\, the main problem is that it is heavily blended with \SiIII. In our emission-line fitting, we use the total flux of the \CIII$+$\SiIII\ complex, same as in \citet{Jiang07}. We do not include narrow components for the \CIII\ complex.

\AlIII\ is relatively weak compared to \CIII\ $+$ \SiIII. It also blends with \CIII\ $+$ \SiIII, but can be easily deblended in most of our objects. We exclude \AlIII\ flux measurements with large uncertainties (more than 30\% of the line flux) or with broader line widths compared to the \CIII\ $+$ \SiIII\ complex. These are due to low S/Ns.

It is sometimes difficult to decompose \HeII\ $\lambda 1640$ and \OIII\ $\lambda 1663$ because of a $\lambda 1600$ feature \citep{Nagao06} that is likely due to imperfect subtraction of \FeII\ and \FeIII\ from the empirical template. In our work, we set the center of \OIII\ and \HeII\ within $[1656,1663]$\,\AA\ and $[1620,1640]$\,\AA, respectively. As for \AlIII, We exclude the flux measurements with large uncertainties.

\SiIV\ is heavily blended with \OIV. We treat \SiIV\ and \OIV\ as one integrated component and use two broad Gaussians to fit. The \SiIV\ line complexes are often weak compared to the flux uncertainties, and suffer from absorption lines. We exclude \SiIV\ measurements with large uncertainties.

Our \NV\ measurements can be affected by the absorption of \NV\ or \lya. We exclude objects that are severely affected by absorption. For the remaining objects, we mask the affected wavelength pixels and fit the line profile with a single Gaussian. In general, our masks provide reasonable fits (see the full figure set of Figure 1 in the online version).

\begin{figure*}[t]

    \includegraphics[width=0.99\textwidth]{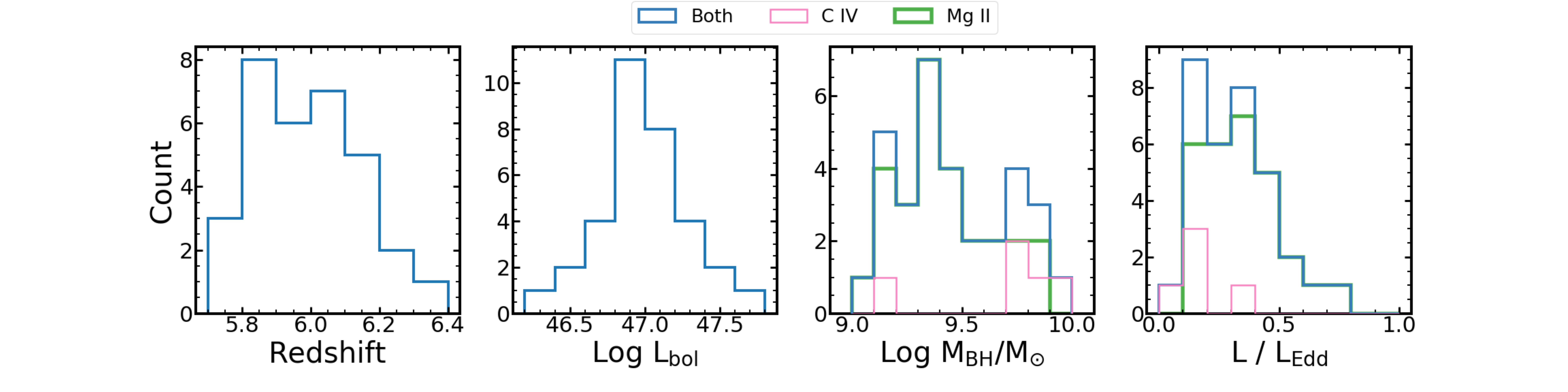}
    \centering
    \caption{Left to right: the distributions of redshift, bolometric luminosity  ${\rm Log}\;{L}_{\rm bol}$, BH mass ${\rm Log}\;{M}_{\rm BH}/{M}_{\rm \odot}$, and Eddington ratio ${L}/{L}_{\rm Edd}$. In the third and fourth panels, the green and pink histograms indicate the the sub-sample where the quantity is derived using \MgII\ and \CIV, respectively, while the blue denotes the sum of the two.}
    \label{fig:AGN_properties_distribution}
\end{figure*}

\subsection{Spectral measurements and derived physical properties}\label{sec:spectral_measurements}

All spectral quantities are measured from the best-fit models. We measure the flux density and monochromatic luminosity at the rest-frame $1350$ and $3000$\,\AA. We also measure the FWHMs of broad \MgII\ and \CIV\ to estimate BH masses. The \FeII\ flux is computed by integrating the best-fit \FeII\ model over the rest-frame wavelength range $2200<\lambda<3090$\,\AA\ \citep{DeRosa11,Shen11}. For broad emission-line flux, we calculate the sum of the multiple Gaussians integrated over the full line profile.

We estimate the uncertainties of our measurements using the Monte Carlo method. We generate 50 mock spectra for each object by adding Gaussian random fluxes at each wavelength pixel to perturb the original spectrum. The Gaussian width is set to the flux uncertainties at each wavelength pixel. The same fitting procedure is applied to all mock spectra to derive the distribution of a given spectral quality. The final uncertainties of spectral measurements are the semi-amplitude of the range enclosing the 16th to 84th percentiles of the distribution.

We present the estimates of quasar BH mass $M_{\rm BH}$, bolometric luminosity $L_{\rm bol}$, and Eddington ratio $L/L_{\rm Edd}$. The BH masses are measured mainly based on the broad \MgII, which are shown to correlate well with those estimated from broad Balmer lines \citep{Wang09,Shen11}. For four objects in our sample that do not have robust \MgII\ measurements, we use the broad \CIV\ line \citep{VestergaardPeterson06,Shen11}. The bolometric luminosity is estimated by $L_{\rm bol} = 5.15 \times \lambda L_{\lambda}$ at $\lambda=3000$\,\AA\  \citep{Richards06}. Uncertainties in these derived quantities are estimated from the error propagation. Table \ref{tab:flux_ratios} lists these properties and their distributions are shown in Figure \ref{fig:AGN_properties_distribution}. Our measurements are generally consistent with those reported in \citet{Shen19a} within uncertainties, with minor differences due to different fitting components (e.g., we added the Balmer continuum component). In particular, our $L_{3000}$ are 0.1 dex lower than those in \citet{Shen19a}, and the \MgII\ FWHM is different in some objects where we included a narrow \MgII\ component. Nevertheless, these minor differences do not impact our results.

%\newpage
{\movetabledown=1.5in 
\begin{deluxetable*}{l c c c c  c c c c c c c}
\rotate 
%\centering % centering table

\tablecaption{Quasar properties and emission line flux ratios}
%\label{tab:flux_ratios}

\tablehead{\colhead{Name} &\colhead{$z_{sys}$} &\colhead{log $L_{\rm bol}$} &\colhead{log $M_{\rm BH}$} &\colhead{Source} &\colhead{$L/L_{\rm Edd}$} &\colhead{\SiIV$/$\CIV} &\colhead{\CIII$/$\CIV} &\colhead{\AlIII$/$\CIV} &\colhead{\HeII$/$\CIV} &\colhead{\NV$/$\CIV} &\colhead{\NV$/$\HeII} }

\footnotesize
%\begin{tabular}{l c c c c  c c c c c c c} %12 cols 
%\hline
\startdata
%Name    &  $z_{sys}$  & $L_{\rm Bol}$ & $M_{\rm BH}$ & Source &  $L/L_{\rm Edd}$ & \SiIV$/$\CIV & \CIII$/$\CIV & \AlIII$/$\CIV & \NV$/$\CIV & \HeII$/$\CIV & \NV$/$\HeII  \\ \hline
J0002$+$2550 & $5.818 \pm 0.007$  & $47.180 \pm 0.003$  & $9.36 \pm 0.05$ & \MgII\  & $0.50 \pm 0.06$  & $0.50 \pm 0.07$  & $0.51 \pm 0.02$  & $0.14 \pm 0.02$  & ...  & ...  & ...  \\ 
J0008$-$0626 & $5.929 \pm 0.006$  & $46.964 \pm 0.013$  & $9.19 \pm 0.07$ & \MgII\  & $0.45 \pm 0.07$  & $0.47 \pm 0.06$  & $0.62 \pm 0.05$  & $0.11 \pm 0.01$  & ...  & $0.99 \pm 0.21$  & ...  \\ 
J0028$+$0457 & $5.982 \pm 0.001$  & $46.969 \pm 0.020$  & $9.91 \pm 0.13$ & \CIV\  & $0.09 \pm 0.02$  & $0.52 \pm 0.07$  & ...  & ...  & ...  & $1.06 \pm 0.29$  & ...  \\ 
J0050$+$3445 & $6.251 \pm 0.006$  & $47.103 \pm 0.007$  & $9.76 \pm 0.12$ & \MgII\  & $0.17 \pm 0.04$  & $0.41 \pm 0.05$  & ...  & ...  & $0.23 \pm 0.06$  & $0.77 \pm 0.15$  & $3.56 \pm 0.15$  \\ 
J0353$+$0104 & $6.057 \pm 0.005$  & $46.975 \pm 0.042$  & $9.32 \pm 0.17$ & \MgII\  & $0.35 \pm 0.10$  & ...  & ...  & ...  & ...  & ...  & ...  \\ 
J0810$+$5105 & $5.805 \pm 0.010$  & $47.193 \pm 0.009$  & $9.29 \pm 0.11$ & \MgII\  & $0.61 \pm 0.14$  & ...  & $0.53 \pm 0.03$  & $0.19 \pm 0.04$  & $0.30 \pm 0.06$  & $1.41 \pm 0.26$  & $4.80 \pm 0.26$  \\ 
J0835$+$3217 & $5.902 \pm 0.009$  & $46.286 \pm 0.006$  & $8.91 \pm 0.10$ & \MgII\  & $0.18 \pm 0.04$  & $0.40 \pm 0.07$  & $0.41 \pm 0.05$  & $0.11 \pm 0.03$  & $0.33 \pm 0.09$  & $0.75 \pm 0.09$  & $2.47 \pm 0.09$  \\ 
J0836$+$0054 & $5.834 \pm 0.007$  & $47.621 \pm 0.004$  & $9.61 \pm 0.08$ & \MgII\  & $0.79 \pm 0.15$  & $0.49 \pm 0.08$  & $0.59 \pm 0.07$  & $0.30 \pm 0.13$  & $0.14 \pm 0.03$  & $1.54 \pm 0.11$  & $11.93 \pm 0.11$  \\ 
J0840$+$5624 & $5.816 \pm 0.010$  & $46.556 \pm 0.013$  & $9.17 \pm 0.12$ & \MgII\  & $0.19 \pm 0.05$  & $0.39 \pm 0.05$  & $0.67 \pm 0.06$  & $0.15 \pm 0.02$  & $0.18 \pm 0.05$  & ...  & ...  \\ 
J0841$+$2905 & $5.954 \pm 0.005$  & $46.986 \pm 0.008$  & $9.40 \pm 0.19$ & \MgII\  & $0.29 \pm 0.12$  & $0.24 \pm 0.06$  & ...  & ...  & ...  & $1.03 \pm 0.22$  & ...  \\ 
J0842$+$1218 & $6.069 \pm 0.009$  & $47.196 \pm 0.005$  & $9.52 \pm 0.06$ & \MgII\  & $0.36 \pm 0.04$  & $0.36 \pm 0.06$  & $0.49 \pm 0.03$  & $0.17 \pm 0.02$  & $0.16 \pm 0.05$  & $0.85 \pm 0.06$  & $5.18 \pm 0.06$  \\ 
J1044$-$0125 & $5.780 \pm 0.007$  & $47.311 \pm 0.006$  & $9.81 \pm 0.10$ & \MgII\  & $0.24 \pm 0.05$  & $0.42 \pm 0.03$  & $0.53 \pm 0.04$  & $0.21 \pm 0.02$  & $0.18 \pm 0.05$  & $1.05 \pm 0.08$  & $5.32 \pm 0.08$  \\ 
J1137$+$3549 & $6.009 \pm 0.010$  & $47.282 \pm 0.008$  & $9.76 \pm 0.09$ & \MgII\  & $0.25 \pm 0.05$  & $0.34 \pm 0.05$  & $0.38 \pm 0.03$  & $0.13 \pm 0.02$  & $0.14 \pm 0.04$  & $0.98 \pm 0.10$  & $5.47 \pm 0.10$  \\ 
J1143$+$3808 & $5.800 \pm 0.010$  & $46.999 \pm 0.005$  & $9.73 \pm 0.08$ & \CIV\  & $0.14 \pm 0.02$  & ...  & $0.52 \pm 0.04$  & $0.07 \pm 0.02$  & $0.21 \pm 0.06$  & $1.00 \pm 0.12$  & $5.09 \pm 0.12$  \\ 
J1148$+$0702 & $6.344 \pm 0.006$  & $47.091 \pm 0.013$  & $9.38 \pm 0.20$ & \MgII\  & $0.39 \pm 0.17$  & $0.46 \pm 0.06$  & $0.46 \pm 0.07$  & $0.17 \pm 0.03$  & $0.20 \pm 0.05$  & $0.75 \pm 0.08$  & $6.04 \pm 0.08$  \\ 
J1148$+$5251 & $6.416 \pm 0.006$  & $47.533 \pm 0.004$  & $9.82 \pm 0.09$ & \MgII\  & $0.40 \pm 0.08$  & $0.31 \pm 0.05$  & ...  & ...  & ...  & ...  & ...  \\ 
J1207$+$0630 & $6.028 \pm 0.013$  & $46.909 \pm 0.011$  & $9.53 \pm 0.08$ & \MgII\  & $0.18 \pm 0.03$  & $0.46 \pm 0.06$  & $0.56 \pm 0.05$  & $0.20 \pm 0.05$  & ...  & ...  & ...  \\ 
J1243$+$2529 & $5.842 \pm 0.006$  & $47.060 \pm 0.004$  & $9.84 \pm 0.05$ & \CIV\  & $0.13 \pm 0.01$  & $0.25 \pm 0.09$  & $0.37 \pm 0.02$  & $0.09 \pm 0.02$  & $0.16 \pm 0.04$  & ...  & ...  \\ 
J1250$+$3130 & $6.138 \pm 0.005$  & $46.988 \pm 0.005$  & $9.13 \pm 0.06$ & \MgII\  & $0.56 \pm 0.07$  & $0.36 \pm 0.02$  & $0.39 \pm 0.03$  & $0.05 \pm 0.03$  & $0.13 \pm 0.03$  & $0.99 \pm 0.07$  & $7.91 \pm 0.07$  \\ 
J1257$+$6349 & $5.992 \pm 0.010$  & $46.739 \pm 0.013$  & $9.43 \pm 0.10$ & \MgII\  & $0.16 \pm 0.03$  & ...  & ...  & ...  & ...  & ...  & ...  \\ 
J1429$+$5447 & $6.119 \pm 0.008$  & $46.831 \pm 0.005$  & $9.18 \pm 0.18$ & \MgII\  & $0.34 \pm 0.14$  & ...  & ...  & ...  & ...  & ...  & ...  \\ 
J1436$+$5007 & $5.809 \pm 0.010$  & $47.044 \pm 0.012$  & $9.30 \pm 0.18$ & \MgII\  & $0.42 \pm 0.16$  & ...  & $0.68 \pm 0.15$  & ...  & ...  & $1.09 \pm 0.17$  & ...  \\ 
J1545$+$6028 & $5.794 \pm 0.007$  & $46.532 \pm 0.013$  & $9.20 \pm 0.05$ & \CIV\  & $0.17 \pm 0.02$  & $0.22 \pm 0.04$  & $0.39 \pm 0.02$  & $0.07 \pm 0.02$  & $0.09 \pm 0.01$  & $1.03 \pm 0.10$  & $11.60 \pm 0.10$  \\ 
J1602$+$4228 & $6.083 \pm 0.005$  & $47.210 \pm 0.009$  & $9.42 \pm 0.08$ & \MgII\  & $0.47 \pm 0.07$  & $0.33 \pm 0.05$  & $0.59 \pm 0.03$  & $0.15 \pm 0.01$  & $0.19 \pm 0.06$  & ...  & ...  \\ 
J1609$+$3041 & $6.146 \pm 0.006$  & $46.645 \pm 0.009$  & $9.44 \pm 0.10$ & \MgII\  & $0.12 \pm 0.02$  & $0.45 \pm 0.09$  & ...  & ...  & $0.16 \pm 0.04$  & $0.92 \pm 0.08$  & $7.89 \pm 0.08$  \\ 
J1623$+$3112 & $6.254 \pm 0.006$  & $46.975 \pm 0.003$  & $9.32 \pm 0.15$ & \MgII\  & $0.35 \pm 0.12$  & $0.33 \pm 0.09$  & $0.46 \pm 0.07$  & $0.06 \pm 0.02$  & $0.12 \pm 0.05$  & $0.91 \pm 0.09$  & $5.89 \pm 0.09$  \\ 
J1630$+$4012 & $6.066 \pm 0.007$  & $46.760 \pm 0.007$  & $9.27 \pm 0.10$ & \MgII\  & $0.24 \pm 0.05$  & $0.17 \pm 0.03$  & $0.27 \pm 0.02$  & $0.06 \pm 0.02$  & ...  & $0.83 \pm 0.13$  & ...  \\ 
J2310$+$1855 & $5.956 \pm 0.011$  & $47.464 \pm 0.005$  & $9.66 \pm 0.15$ & \MgII\  & $0.49 \pm 0.16$  & $0.46 \pm 0.05$  & ...  & ...  & ...  & ...  & ...  \\ 
P000$+$26 & $5.733 \pm 0.007$  & $47.333 \pm 0.034$  & $9.70 \pm 0.09$ & \CIV\  & $0.33 \pm 0.04$  & ...  & $0.55 \pm 0.04$  & $0.06 \pm 0.03$  & $0.16 \pm 0.03$  & ...  & ...  \\ 
P060$+$24 & $6.170 \pm 0.006$  & $47.057 \pm 0.013$  & $9.32 \pm 0.03$ & \MgII\  & $0.42 \pm 0.02$  & $0.47 \pm 0.07$  & $0.54 \pm 0.09$  & $0.12 \pm 0.02$  & $0.17 \pm 0.05$  & $1.10 \pm 0.19$  & $6.51 \pm 0.19$  \\ 
P210$+$27 & $6.166 \pm 0.007$  & $46.910 \pm 0.017$  & $9.33 \pm 0.11$ & \MgII\  & $0.29 \pm 0.06$  & $0.35 \pm 0.09$  & ...  & $0.11 \pm 0.03$  & $0.13 \pm 0.03$  & ...  & ...  \\ 
P228$+$21 & $5.893 \pm 0.015$  & $46.605 \pm 0.007$  & $9.07 \pm 0.06$ & \MgII\  & $0.26 \pm 0.03$  & ...  & $0.57 \pm 0.06$  & ...  & ...  & ...  & ...  \\ 
P333$+$26 & $6.027 \pm 0.007$  & $46.839 \pm 0.017$  & $9.24 \pm 0.10$ & \MgII\  & $0.30 \pm 0.05$  & ...  & $0.57 \pm 0.13$  & $0.17 \pm 0.05$  & $0.18 \pm 0.06$  & ...  & ...  \\    \hline
Median &  $5.992\pm0.172$  & $46.990\pm0.289$ &  $9.36\pm0.25$ & All &  $0.30\pm0.16$  &  $0.39^{+0.07}_{-0.10}$ & $0.53^{+0.06}_{-0.13}$ &  $0.12^{+0.06}_{-0.06}$& $0.17^{+0.04}_{-0.03}$  &  $0.99^{+0.10}_{-0.17}$ &  $5.67^{+2.23}_{-0.86}$ \\
Composite &  ...  & ... &  ... & ... &  ...  &  $0.33\pm0.06$ & $0.48\pm0.02$ &  $0.13\pm0.02$& $0.15\pm0.02$  &  $0.95\pm0.15$ &  $5.70\pm0.65$ \\
\enddata
%\end{tabular}
\end{deluxetable*} 

}\label{tab:flux_ratios}
%\newpage

\section{Results}
\subsection{UV line flux ratios and metallicity} \label{sec:uvratio_metallicity}

Table \ref{tab:flux_ratios} summarizes the results of the main diagnostic line flux ratios for metallicity estimates, including \SiIV$/$\CIV, \CIII$/$\CIV, \AlIII$/$\CIV, \NV$/$\CIV, \HeII$/$\CIV, and \NV$/$\HeII. In this section we explore the dependence of these flux ratios on redshift, and estimate BLR metallicities using empirical relations calibrated by photoionization models. For simplicity, we use the name of the main line in each line complex to refer to the whole line complex, e.g., \SiIV\ for \SiIV$+$\OIV and \CIII\ for \CIII$+$\SiIII. 

\begin{figure*}[htbp]
   \centering
    \includegraphics[width=0.99\textwidth]{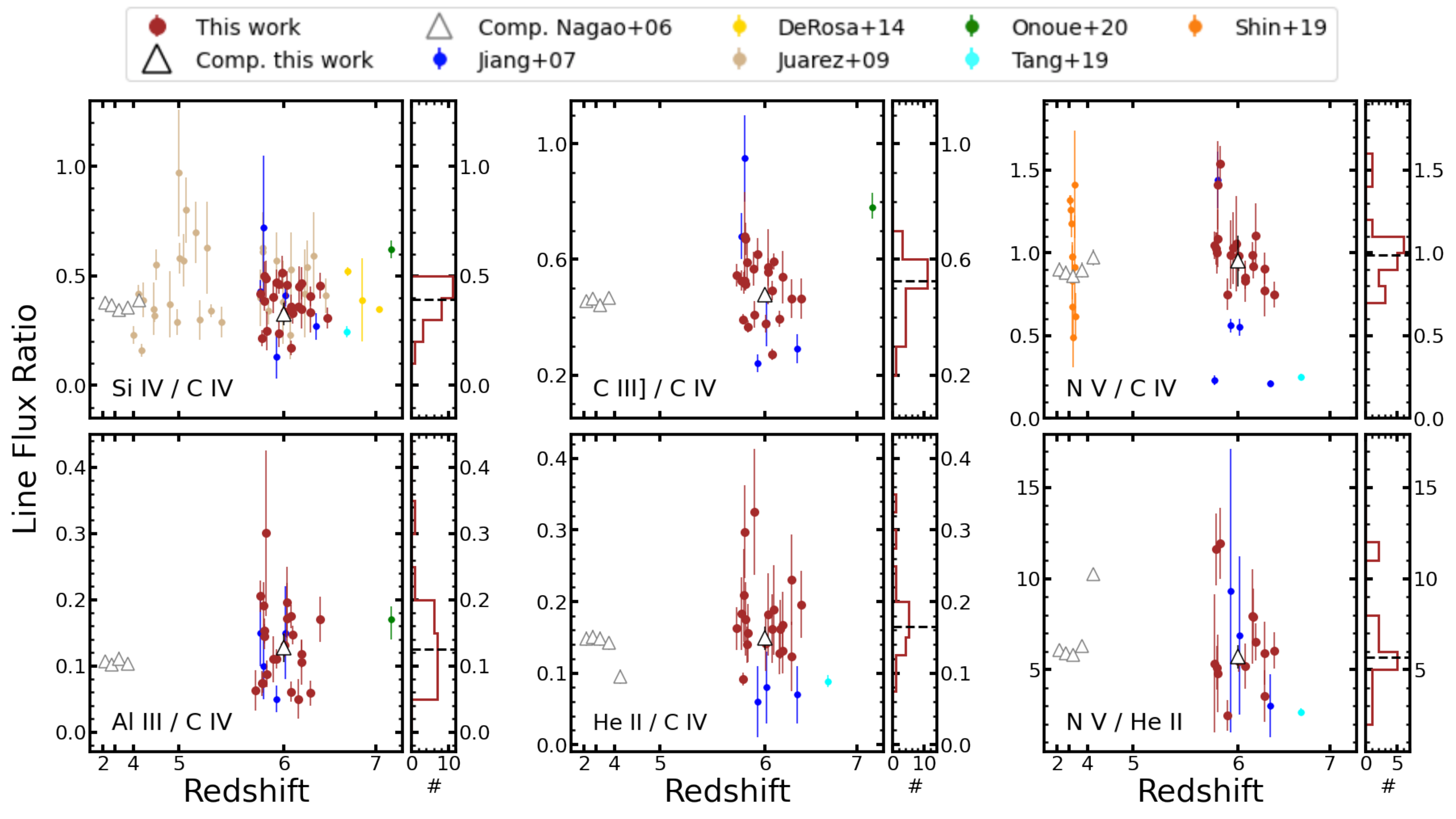} 
    \caption{Different line flux ratios (labeled at the lower left corner) as a function of redshift. Different samples are plotted with different colors, as shown in the top legend. The points represent individual measurements, while the triangles are the values measured from composite spectra. The x-axis is scaled for display purpose. For each panel of flux ratios, we plot the histogram of the line flux ratio distribution for our sample on the right sidebar. The black dashed horizontal lines indicate the median values of our sample. 
 }
   \label{fig:lineratios1}
\end{figure*}

\begin{table*}[htbp]

    \caption{Sample Summary}
    \label{tab:sample_summary}
    \tabcolsep 6pt
    \centering
    \small
    \begin{tabular}{l c c c c c}
    \hline \hline
     Sample            &  Redshift & Median $M_{B}$ & $M_B$ range    &  Size  &  Metallicity indicator$^a$\\ \hline
     
     This work            &   $5.7<z<6.4$  &  $-27.2$ & $-26.4\sim-28.5$  & 30$^b$ & \SiIV$/$\CIV, \CIII$/$\CIV, \AlIII$/$\CIV \\
      & & &  &  & \HeII$/$\CIV, \NV$/$\CIV,  \NV$/$\HeII\ \\ 
      
     \multicolumn{6}{c}{Other high redshift sample } \\ 
     
    \citet{Onoue20}     &   $7.5$     &  $-27.2$ & ... &  1 & \SiIV$/$\CIV, \CIII$/$\CIV, \AlIII$/$\CIV \\ 
    
    \citet{DeRosa14}     &   $6.6<z<7.1$   &  $-26.4$ & $-26.1\sim-27.2$   &  4 & \SiIV$/$\CIV\ \\
    
    \citet{Tang19}   &   6.6     &    $-26.1$ & ... &  1 & \SiIV$/$\CIV,  \HeII$/$\CIV, \NV$/$\CIV, \NV$/$\HeII \\
    
    \citet{Jiang07}    &   $5.8<z<6.3$     &  $-27.6$ & $-27.2\sim-28.5$   &  6 & Same as this paper \\  
    
    \citet{Juarez09}     &   $4<z<6.4$  &  $-27.2$ &  $-26.2\sim-29.0$    &  30 & \SiIV$/$\CIV\ \\

    \multicolumn{6}{c}{Lower redshift sample } \\ 
    
    \citet{Nagao06}    &   $2.0<z<4.5$ & ... & $-26.5\sim-28.5$ &  2317 & Same as this work\\
    \citet{Shin19}     &   $3.0<z<3.4$     &  $-25.9$ & $-24.7\sim-28.7$    &  12 & \NV$/$\CIV\ \\ \hline
     
     \multicolumn{6}{p{0.95\textwidth}}{Note. a. \citet{DeRosa14} and \citet{Tang19} fit \CIII$+$\SiIII$+$\AlIII\ as a single line and we do not include these measurements here. The \NV$/$\HeII\ ratio in \citet{Jiang07} was estimated from \NV$/$\CIV\ and \HeII$/$\CIV. b. Three objects in our sample cannot obtain robust fit for \CIV, thus cannot be used to study metallicity. They are only used in the study of \FeII$/$\MgII.}
    \end{tabular}

\end{table*}

To study redshift evolution, we compare our sample ($5.7<z<6.5$) with earlier samples at other redshifts ($2<z<7.5$) \citep{Nagao06,Jiang07,Juarez09,DeRosa14,Shin19,Onoue20}. \citet{Nagao06} studied quasar flux ratios at $2<z<4.5$ and $-29.5<{M}_{B}<-24.5$. As mentioned in the introduction, a significant correlation between metallicity and $M_{B}$ was found. Therefore, to study the redshift evolution, we need to compare samples with similar luminosities. The $M_B$ magnitudes of our sample are derived from $M_{1450}$ compiled from the literature \citep{Jiang16,Willott10a,Banados16}, assuming a spectral index $\alpha_{\upsilon}=-0.5$ following \citet{Richards06}. Note that for our quasars, the conversion of $M_B$ from $M_{1450}$ is an extrapolation from the GNIRS spectrum, but the adopted power-law index is a good approximation in the range of $1450<\lambda_{\rm rest}<5000$\,\AA\ \citep{Vandenberk01}. The median $M_B$ value of our sample is $-27.2$, ranging from $-26.4$ to $-28.5$. Therefore, we select two bins from \citet{Nagao06}, $-26.5<{M}_{B}<-27.5$ and $-27.5<{M}_{B}<-28.5$, and calculated average flux ratios in $-26.5<{M}_{B}<-28.5$ in each redshift bin weighted by the number of objects in each bin. Apart from \citet{Nagao06}, other studies also provide various line flux ratios in different redshift ranges \citep{Jiang07,Juarez09,Shin19,Onoue20}. We summarize these earlier samples in Table~\ref{tab:sample_summary}.

Figure \ref{fig:lineratios1} displays various line ratios as a function of redshift compiled from different samples. Although the scatter is large, the median values of these line flux ratios are consistent with no redshift evolution\footnote{Note that the  $4.0<z<4.5$ redshift bin of \citet{Nagao06} deviates from the average trends in some ratios, e.g., \HeII$/$\CIV, \NV$/$\HeII, which is noted by \citet{Nagao06} as well. It could be due to the small sample size in their $-28.5<M_{B}<-27.5$ and selection biases.} within $2\lesssim z\lesssim 6$. These median values of our sample are summarized in Table \ref{tab:flux_ratios}. We also generate a high S/N median composite spectrum of our final sample following \citet{Shen19a}, and found consistent results in the average line flux ratios (open triangles in Fig.~\ref{fig:lineratios1}).

Our results are generally consistent with other high-redshift studies \citep{Jiang07,Juarez09,DeRosa14,Tang19,Onoue20}. Two objects, J0836$+$0054 and J1044$-$0125, overlap with both \citet{Jiang07} and \citet{Juarez09}. Additional five objects, J0002$+$2550, J1148$+$5251, J1602$+$4228, J1623$+$3112, and J1630$+$4012, overlap with \citet{Juarez09}. The \SiIV$/$\CIV\ ratios of these objects are consistent ($\le1\sigma$) among different studies. For \NV$/$\CIV\ in the two objects overlapping with \citet{Jiang07}, one object is consistent within 1$\sigma$ uncertainty and the other one shows a difference. It is likely due to the different spectral fitting procedures. For example, \citet{Jiang07} fixed the \NV\ line centroid to the value given in \citet{Vandenberk01}, while in our work it is allowed to vary in the range of $[1230,1240]$\,\AA. We find that the \NV\ centroid of our best-fit models is usually blue-shifted. Another difference resides in the \AlIII$/$\CIV\ and \CIII$/$\CIV line ratios. The two overlapped objects in \citet{Jiang07} have larger \CIII$/$\CIV\ and lower \AlIII$/$\CIV\ ratios, which is resulted from the different decomposition strategies. Given the small overlap sample, these differences are not significant. Table \ref{tab:sample_summary} shows that our work substantially increases the sample size at $z\sim6$ with multiple metallicity diagnostic ratios measured.

Utilizing the average flux ratios measured from the composite spectrum of our sample ($z\sim6$) and those at lower redshifts \citep{Nagao06}, we examine the correlation between different line flux ratios and redshift using Pearson's correlation test. The Pearson correlation coefficients $r$ and the null-hypothesis significance $p(r)$ are presented in Table \ref{tab:pearson_r_redshift}. In summary, we find no significant redshift evolution in these metallicity diagnostic ratios up to $z\sim6$. We also measure the slope in the flux ratio-redshift relation using the Bayesian linear regression package {\tt Linmix} \citep{Kelly07}. The slopes are consistent with zero within uncertainties. 

\begin{table}[b]

    \caption{Slopes from the linear regression fits, and the Pearson correlation coefficients $r$ and significance $p(r)$ of the correlations between redshift and line flux ratios measured from several composite spectra at different redshift (\citet{Nagao06} and this work).} \label{tab:pearson_r_redshift}
    \small
    \centering
    \tabcolsep 8pt
    \begin{tabular}{c  c  c c }
    \hline \hline
    Indicator   &  slope            &$r$  &  $p(r)$ \\ \hline
    \SiIV$/$\CIV  &  $-0.006\pm0.019$ & $-0.55$  &  $0.253$ \\ 
    \CIII$/$\CIV  &  $+0.007\pm0.019$ & $+0.69$  &  $0.196$ \\ 
    \AlIII$/$\CIV &  $+0.006\pm0.013$ & $+0.85$  &  $0.067$ \\ 
    \NV$/$\CIV    &  $+0.026\pm0.039$ & $+0.69$  &  $0.131$ \\ 
    \HeII$/$\CIV  &  $-0.000\pm0.010$ & $-0.02$  &  $0.969$ \\ 
    \NV$/$\HeII   &  $+0.246\pm1.178$ & $+0.15$  &  $0.778$ \\ \hline
    \end{tabular}

\end{table}

\begin{figure*}[t]
    \centering
    \includegraphics[width=0.99\textwidth]{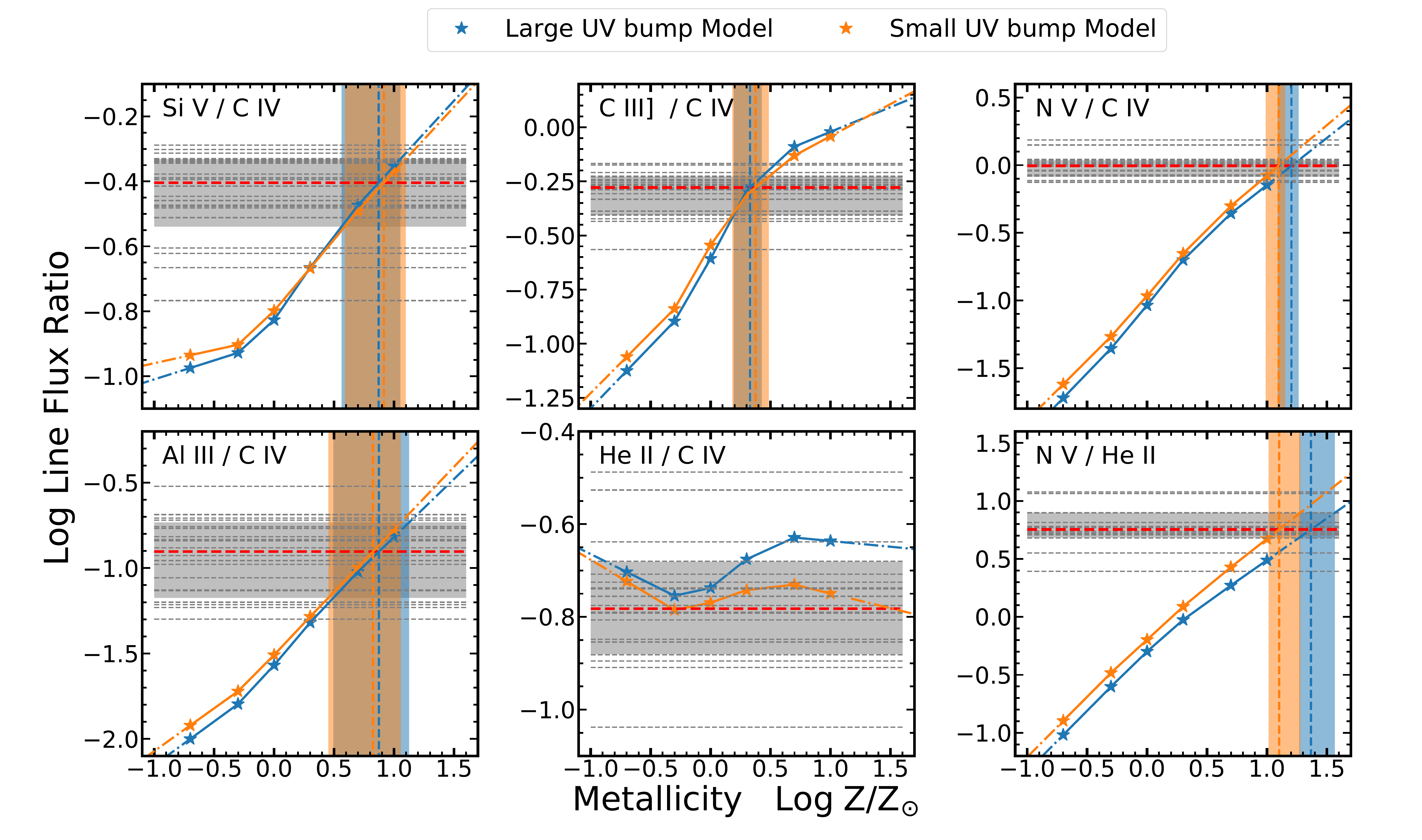}
    \caption{Illustration of converting line ratios to metallicities using photoionization models \citep{Nagao06}. The blue and orange asterisks (and solid lines) represent the model predictions (and linear interpolation) from SEDs with large and small UV bumps, respectively. The dash-dotted lines are the extrapolation of these models. Each grey horizontal dashed line represents individual object in our sample, and red dashed lines represent the median ratio. The horizontal grey shaded areas are the range between 16th and 84th percentile of the line flux ratio distribution. The blue and orange vertical lines are the median metallicities indicated from the large and small UV bump models, respectively. The blue and orange shaded areas represent the range between 16th and 84th percentiles of the inferred metallicity distribution. }
    \label{fig:metallicity}
\end{figure*}

Next, we convert the line flux ratios to metallicities. We compare the median value of each line flux ratio with the photoionization predictions from the locally optimally emitting cloud (LOC) model \citep{Baldwin95}. \citet{Hamann02} and \citet{Nagao06} carried out detailed simulations based on CLOUDY \citep{Ferland98} and studied how line flux ratios vary as a function of BLR metallicity. We utilize the model predictions in Table 10 of \citet{Nagao06}.  Both small and large UV bump SED models are considered in our work.

Specifically, \citet{Nagao06} provides the model predictions at ${\rm Z}/{\rm Z}_{\odot}=0.2, 0.5, 1.0, 2.0, 5.0, 10.0,$ for \NV$/$\CIV, \SiIV$/$\CIV, \CIII$/$\CIV, \AlIII$/$\CIV, and \HeII$/$\CIV. The \NV$/$\HeII\ ratio can be calculated using \NV$/$\CIV\ and \HeII$/$\CIV. Figure \ref{fig:metallicity} illustrates how we estimate the metallicity from individual line ratios. We linearly interpolate (and extrapolate if necessary) the model predicted metallicity-line flux ratio curves and convert the line ratios to metallicity from the interpolation (extrapolation). Table \ref{tab:metallicity} summarizes the estimated metallicities. We fail to convert \HeII$/$\CIV\ to metallicity because most of the measured \HeII$/$\CIV\ ratios do not overlap with the model predictions, which is also noted by \citet{Nagao06}.

\begin{table}[h]

  \caption{BLR metallicity of our sample estimated from median line flux ratios or flux ratios measured from the composite spectrum, using photoionization model predictions \citep{Nagao06}. } \label{tab:metallicity}
    \small
    \centering
    \tabcolsep 8pt
    \begin{tabular}{c  r r}
    \hline \hline
       Indicator       &        \multicolumn{2}{c}{Metallicity ($Z/Z_{\odot}$)} \\ 
                       & Large UV bump & Small UV bump\\ \hline
      \multicolumn{3}{c}{Median Metallicity of our sample} \\
      \SiIV$/$\CIV  & $7.4_{-3.8}^{+3.9}$  & $8.2_{-4.4}^{+4.3}$  \\
      \CIII$/$\CIV  & $2.1_{-0.6}^{+0.5}$  & $2.4_{-0.9}^{+0.7}$  \\
      \AlIII$/$\CIV  & $7.5_{-4.4}^{+5.9}$  & $6.8_{-3.8}^{+4.7}$  \\  
      \NV$/$\CIV    & $16.0_{-4.0}^{+2.3}$  & $12.5_{-3.0}^{+1.7}$  \\
      \NV$/$\HeII    & $23.3_{-4.7}^{+13.6}$  & $12.6_{-2.3}^{+6.4}$  \\

   \multicolumn{3}{c}{Metallicity from the composite spectrum} \\
    \SiIV$/$\CIV  & ${4.7}_{-1.6}^{+1.8}$  & ${5.1}_{-1.8}^{+2.0}$  \\ 
    \CIII$/$\CIV & ${1.9}_{-0.1}^{+0.1}$  & ${2.0}_{-0.1}^{+0.1}$  \\ 
     \AlIII$/$\CIV & ${7.7}_{-2.0}^{+2.0}$  & ${6.8}_{-1.6}^{+1.6}$  \\ 
     \NV$/$\CIV    & ${15.1}_{-3.5}^{+3.5}$  & ${11.9}_{-2.6}^{+2.6}$  \\ 
    \NV$/$\HeII    & ${23.4}_{-3.7}^{+3.7}$  & ${12.7}_{-1.8}^{+1.8}$  \\

      \hline
    \end{tabular}
 
\end{table}

\begin{figure*}[t]
    \centering
    \includegraphics[width=0.99\textwidth]{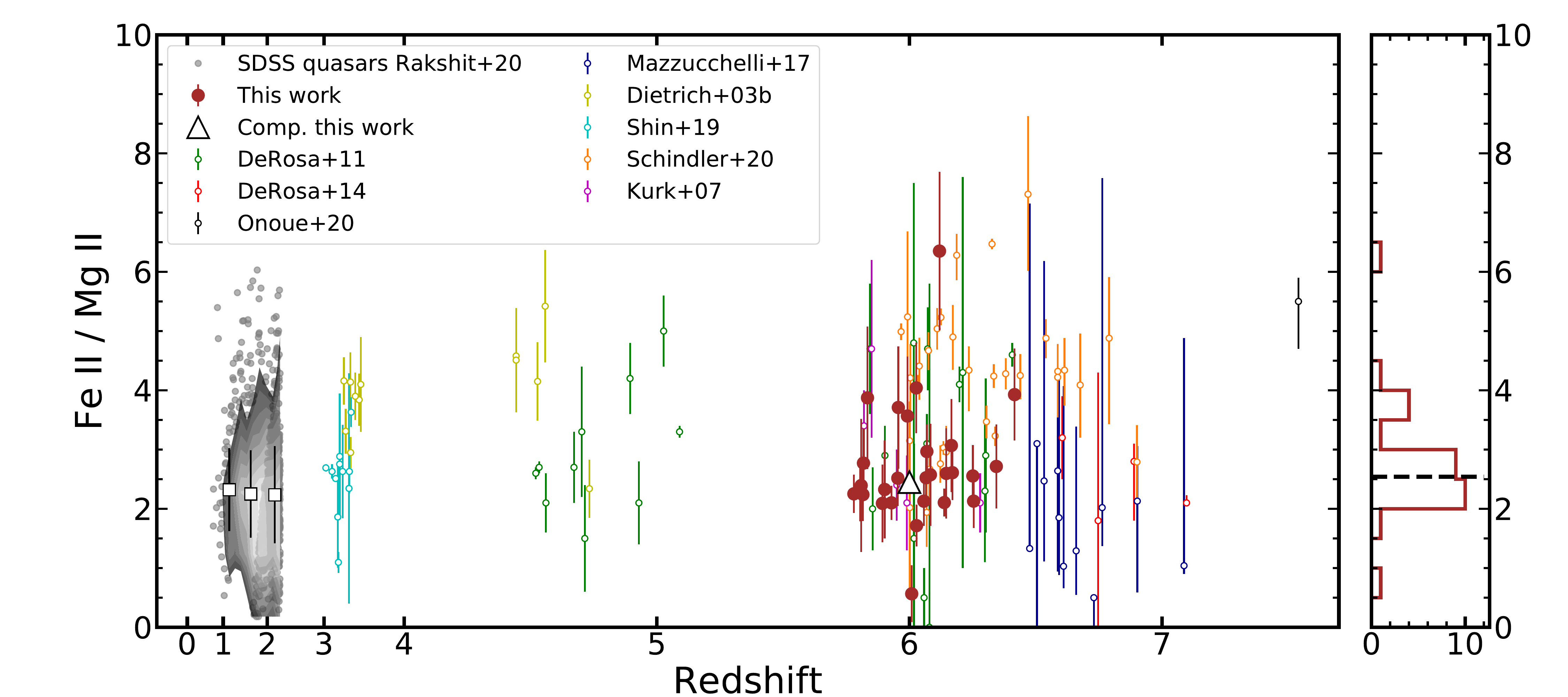}
    \caption{\FeII$/$\MgII\ ratios as a function of redshift. Different samples are indicated by different colors presented in the legend. The points represent individual measurement, while the triangles refer to the value of composite spectrum. The shaded area and the contours indicate the distribution of luminosity distribution matched low redshift SDSS quasars \citep{Rakshit20}, and the squares refer to the median values in different redshift bins. We also display the histogram of \FeII$/$\MgII\ ratio distribution of our sample at the right panel, and the black dashed line refers to the median value. The x-axis is  scaled for display purpose.}
    \label{fig:fe2mg2_vs_z}
\end{figure*}

Converting line ratios to metallicities is subject to considerable systematic uncertainties. The conversion can be affected by factors such as the ionizing parameter, the hardness of the ionizing continuum, temperature, density, etc. (see \citet{Maiolino19} and reference therein). It is proposed that the relative abundance of N element can be used as metallicity diagnostics  \citep[e.g.,][]{Hamann02}, including \NIII$/$\OIIIUV,  \NV/(\CIV+\OVI), \NV/\CIV, and \NV/\HeII. However, the line ratios involving the N element may be more related to N over-abundance rather than metallicity \citep{Jiang08}. In addition, the line ratios that involve major UV lines, i.e., \NV/\CIV\ and \NV/\HeII, are sensitive to ionization parameter and shape of the ionizing continuum \citep{Maiolino19}. Our results support this idea, as the inferred metallicities, ${\rm Z}_{\rm N\,{\rm  V}/C\,{\rm IV}}$ and ${\rm Z}_{\rm N\,V/He\,II}$, are much higher than those from other indicators. On the other hand, this could also be a consequence of possible biases in our fitting, since \lya\ often cannot be well modeled because of absorption in high-redshift quasars. But the median values of \NV$/$\CIV\ and \NV$/$\HeII\ agree with those measured from lower redshifts, suggesting that the \lya\ absorption did not largely bias our \NV\ measurements.  Other preferred metallicity indicators are \SiIV$/$\CIV\ and \AlIII$/$\CIV. These indicators do not involve the N element and  are suggested to be more robust \citep{Maiolino19} since they are not sensitive to differences in the ionizing continuum \citep{Nagao06, Matsouka11, Maiolino19}. As for \CIII$/$\CIV, our results are similar to \citet{Nagao06}, who found $Z<2{Z_{\odot}}$ using \CIII$/$\CIV, reflecting the broad range of inferred metallicities from multiple line ratios using photoionization calculations.

In this work, we will not try to understand the discrepancy between each metallicity indicator, and neither will we draw any conclusion about which one should be adopted as the real metallicity. While individual line flux ratios do not provide accurate BLR metallicity estimates, we use the average metallicity over multiple line ratios to conclude that: 1) the BLR metallicity is at least supersolar even at $z\sim6$; 2) for each of the metallicity indicators, there is no obvious trend with redshift  up to $z\sim6$. Our results are in agreement with the metallicity measurements in \citet{Jiang07}, who obtained metallicity estimate of $4Z_{\odot}$ at $z\sim6$. These results suggest that the quasar BLRs are already metal enriched at $z\sim6$, and the enrichment of the BLR metallicity must occur at earlier times and/or very quickly.

\begin{table}

\caption{\FeII$/$\MgII\ ratios using VW01 and T06 template}\label{tab:fe2tomg2_ratios}
\doublerulesep 0.1pt \tabcolsep 13pt
\small
\centering 
\begin{tabular}{l c c} 
\hline
Name    &  \multicolumn{2}{c}{\FeII$/$\MgII} \\
        &  VW01 & T06 \\  \hline
        
J0002$+$2550 & $2.77 \pm 0.59$  & $4.29 \pm 1.11$  \\ 
J0008$-$0626 & $2.10 \pm 0.29$  & $3.81 \pm 0.45$  \\ 
J0050$+$3445 & $2.55 \pm 0.52$  & $4.22 \pm 0.98$  \\ 
J0353$+$0104 & $2.13 \pm 0.42$  & $3.02 \pm 0.62$  \\ 
J0810$+$5105 & $2.33 \pm 0.54$  & $3.64 \pm 0.92$  \\ 
J0835$+$3217 & $2.33 \pm 0.82$  & $3.99 \pm 1.24$  \\ 
J0836$+$0054 & $3.87 \pm 1.20$  & $9.08 \pm 1.84$  \\ 
J0840$+$5624 & $2.24 \pm 0.44$  & $4.02 \pm 1.15$  \\ 
J0841$+$2905 & $2.52 \pm 0.47$  & $3.95 \pm 0.81$  \\ 
J0842$+$1218 & $2.97 \pm 0.46$  & $5.06 \pm 0.95$  \\ 
J1044$-$0125 & $2.25 \pm 0.33$  & $3.59 \pm 0.96$  \\ 
J1137$+$3549 & $0.57 \pm 0.48$  & $0.74 \pm 0.61$  \\ 
J1148$+$0702 & $2.72 \pm 0.71$  & $5.51 \pm 1.00$  \\ 
J1148$+$5251 & $3.93 \pm 0.78$  & $6.97 \pm 1.22$  \\ 
J1207$+$0630 & $1.72 \pm 0.35$  & $3.08 \pm 0.67$  \\ 
J1250$+$3130 & $2.11 \pm 0.23$  & $4.31 \pm 0.47$  \\ 
J1257$+$6349 & $3.57 \pm 1.00$  & $5.31 \pm 1.72$  \\ 
J1429$+$5447 & $6.35 \pm 1.34$  & $13.12 \pm 2.89$  \\ 
J1436$+$5007 & $2.39 \pm 1.12$  & $2.78 \pm 1.86$  \\ 
J1602$+$4228 & $2.57 \pm 0.86$  & $3.92 \pm 1.03$  \\ 
J1609$+$3041 & $2.60 \pm 0.76$  & $4.21 \pm 1.12$  \\ 
J1623$+$3112 & $2.13 \pm 0.45$  & $3.05 \pm 0.80$  \\ 
J1630$+$4012 & $2.52 \pm 0.46$  & $3.29 \pm 0.89$  \\ 
J2310$+$1855 & $3.71 \pm 1.03$  & $5.30 \pm 1.84$  \\ 
P060+$2$4 & $2.61 \pm 0.46$  & $4.14 \pm 0.87$  \\ 
P210+$2$7 & $3.07 \pm 0.79$  & $4.29 \pm 1.38$  \\ 
P228+$2$1 & $2.09 \pm 0.66$  & $3.09 \pm 1.21$  \\ 
P333+$2$6 & $4.04 \pm 0.77$  & $6.10 \pm 1.39$  \\ 
\hline
Median  &  $2.54^{+1.12}_{-0.43}$ &  $4.08^{+1.36}_{-1.00}$   \\
Composite & $2.45 \pm 0.05$  &   $4.31 \pm 0.11$   \\
\hline
\end{tabular}
\label{tab:LPer}

\end{table}

\subsection{\FeII$/$\MgII}

Core collapsed supernovae eject a comparable mass of Fe and Mg elements \citep[e.g.][]{Tsujimoto95,Nomoto97b} and produce a baseline [Fe/Mg] of roughly $-0.7\sim-0.4$ \cite[][and reference therein]{Hamann99}. On the other hand, SN Ia produce a larger amount of Fe relative to Mg and other $\alpha$ elements \citep[e.g.][]{Tsujimoto95, Nomoto97a}. A rapid increase of the Fe$/$Mg abundance ratio is thus expected at the time when the progenitors of SN Ia began to explode \citep{Venkatesan04}. The commonly quoted timescale of SN Ia, $t_{\rm Ia}$, is $\sim1$ Gyr \citep{Yoshii96}, which is close to the age of the universe at $z\sim6.0$. Therefore, the \FeII$/$\MgII, used as the first order approximation of the Fe$/$Mg abundance ratio, can put important constraints on the star formation history at the early universe.

Figure \ref{fig:fe2mg2_vs_z} compares the \FeII$/$\MgII\ ratios of our sample with other studies from $z\sim0.7$ to $z\sim7.5$ \citep{Dietrich03,Kurk07,DeRosa11,DeRosa14,Mazzucchelli17,Shin19,Schindler20,Onoue20,Rakshit20}. For individual measurements, we only include literature samples that used the same \FeII\ template (VW01) and the same \FeII\ flux integration window ($2200<
\lambda{\rm rest}<3090$\,\AA) as in the current work. In addition, we supplement a sample of low-redshift SDSS quasars from Data Release 14, whose spectral properties are measured by \citet{Rakshit20}\footnote{\citet{Rakshit20} performed their spectral fitting based on {\tt PYQSOFIT} developed by \citet{Guo18}, which is similar to our work. However, they used a modified \FeII\ template \citep{Shen19b} that augments the VW01 template with the \citet{Salviander07} \FeII\ template in $2200<\lambda_{\rm rest}<3090$\,\AA, and the T06 template in $3090<\lambda_{\rm rest}<3500$\,\AA. The difference in \FeII$/$\MgII\ using this template and the VW01 template is $\sim$5\%, much smaller compared to that between the T06 and VW01 templates (see Figure 1 in \citet{Yu21}). }. We select these SDSS quasars at $0.7<z<2.6$ that have similar luminosities as our sample quasars. We further require their spectral S/N$>10$. In addition to the luminosity-matched control sample, we also select SDSS quasars with a matched $L/L_{\rm Edd}$ as our Gemini sample. It shows that there is no obvious difference in the \FeII$/$\MgII\ distribution between luminosity-matched sample and $L/L_{\rm Edd}$-matched sample for the SDSS quasars. We thus use the luminosity-matched SDSS quasars as our main low-redshift control sample. There are also earlier studies that provided the \FeII$/$\MgII\ ratios using the T06 \FeII\ templates. We present the comparison using the T06 template in \ref{sec:appendix_1}. The measurements using both VW01 and T06 templates are summarized in Table \ref{tab:fe2tomg2_ratios}. 

As shown in Figure \ref{fig:fe2mg2_vs_z} and Table \ref{tab:fe2tomg2_ratios}, the median \FeII$/$\MgII\ ratio of our sample ($2.54^{+1.12}_{-0.43}$) is consistent with measurements at low redshifts, e.g., low-redshift SDSS quasars and other samples. The median values of SDSS quasars in three redshift bins at $z\sim1.20$, 1.68, and 2.16 are $2.32\pm0.64$, $2.29\pm0.72$, and $2.24\pm0.82$, respectively. They are all consistent with our median value within $1\sigma$ uncertainty. The \FeII$/$\MgII\ ratio measured from our composite spectrum ($2.45\pm0.05$) also confirms no evolution up to $z\sim 6$. Our results are also generally consistent with other studies for high-redshift quasars \citep{Kurk07,DeRosa11,DeRosa14,Mazzucchelli17, Schindler20}. 

There are two quasars in our sample, J1429+5447 and J1257+6349, labeled as weak-line quasars (WLQ) by \citet{Shen19a}. Their \FeII$/$\MgII\ ratios are 6.35 and 3.57, respectively. In their spectra, \CIV\ is barely detected. Their equivalent widths are less than 10\,\AA, while \MgII\ is detected. On the other hand, strong UV \FeII\ emission is present, which is a typical feature of WLQs \citep{Wu11}. WLQs may have an intrinsically different \FeII$/$\MgII\ distribution and thus complicate the interpretation. Indeed, J1429+5447 has the highest \FeII$/$\MgII\ ratio in our sample, while J1257+6349 is marginally within the 16th$-$84th range. If we exclude these two WLQs, the median \FeII$/$\MgII\ ratio decreases from 2.54 to 2.52. Therefore, the two WLQs in our sample have a negligible effect on our conclusions. 

The Fe$/$Mg element abundance ratio is not the only factor that affects the \FeII$/$\MgII\ ratio. Detailed photoionization calculations suggest that the \FeII$/$\MgII\ ratio has a strong dependence on the velocity of microturbulence \citep[e.g.][]{Verner03,Baldwin04, Sarkar21, Panda21}. Though the preferred turbulence velocity is different among different studies, the existence of microturbulence is found to be the key to reproduce the observed strength and shape of the \FeII\ UV bump \citep{Verner03, Baldwin04, Sarkar21}. The calculated \FeII$/$\MgII\ ratio varies with the turbulence velocity: one dex difference in the turbulence velocity will cause $0.3\sim0.5$ dex difference in \FeII$/$\MgII\ \citep{Verner03}. On the other hand, one dex difference in the Fe abundance will result in $\sim0.3$ dex difference in \FeII$/$\MgII\ \citep{Verner03}. Even if the affect of microturbulence is determined, \FeII$/$\MgII\ also correlates with the $L/L_{\rm Edd}$ \citep[e.g.][]{Dong09, Sameshima17, Shin21}.  All of these largely complicates the interpretation of the \FeII$/$\MgII\ ratio.  A more detailed and consistent study that determines the effect of the microturbulence is needed to fully understand the problem.

If we assume that the \FeII$/$\MgII\ line flux ratio reflects the Fe$/\alpha$ abundance ratio, the lack of redshift evolution of \FeII$/$\MgII\ at $z\sim 6$ (and even $z\sim7$, combined with other studies) seems to challenge the commonly quoted timescale of SN Ia $t_{\rm Ia}$. The timescale $t_{\rm Ia}$ can vary with star formation history \citep[e.g.,][]{Matteucci01}.  In the case of elliptical galaxies in which star formation was initially very efficient and stopped after a short duration ($<0.4$ Gyr),  $t_{\rm Ia}$ can be as short as $0.3$ Gyr. For an instantaneous star formation, $t_{\rm Ia}$ is only $\sim40$ Myr. Observational studies of SN Ia delay time (same as $t_{\rm Ia}$) distribution also found that the rate of SN Ia follows the form of $t^{-1}$ and the highest SN Ia rate occurs at the time as short as 0.2 Gyr after the initial star burst \citep[e.g.,][]{Totani08,Maoz12,Maoz14,Wiseman21}. Therefore, the lack of the \FeII$/$\MgII\  evolution at $z\sim6$ suggests that the initial star formation happened at least before  $z\sim8.1$ ($t_{\rm Ia}\sim0.3$ Gyr) or $z\sim7.3$ ($t_{\rm Ia}\sim0.2$ Gyr) for these quasars.

Finally, we discuss several additional factors that could bias our measurements and/or produce the scatter seen in our measurements. In addition to the choice of the \FeII\ template and the \FeII\ flux integration window, other possible factors include the fitting range, inclusion of narrow \MgII\ components, etc. We use the composite spectrum to fit the $2200<\lambda_{\rm rest}<3300$\,\AA\ range instead of the windows adopted in our fiducial spectral analysis (\S\ref{sec:continuum_model}), because in many studies their spectra only cover this wavelength range. We find that the derived \FeII$/$\MgII\ ratio is 2.85 if adopting the new fitting windows. We also perform a test using a single broad Gaussian for \MgII, which only changes the results by less than 5\%. Therefore, we conclude that the uncertainties in our fitting methodology do not bias our conclusions.

\section{Discussion}

We do not find an apparent redshift evolution of the quasar BLR metallicity at redshift up to $z\sim6$. This is in contrast to star-forming galaxies and DLAs in which a strong redshift evolution of metallicity at $0<z<3.5$ has been firmly established  \citep[e.g.,][]{Maiolino08,Mannucci09}. To explain this difference, many studies attribute the non-evolution in quasars to selection biases \citep{Juarez09,Maiolino19}, such that only the most massive quasars at high redshift are observed. Combined with the BH mass-BLR metallicity relation discovered at lower redshifts \citep{Matsouka11,Xu18}, we are selectively observing quasar BLRs with high metallicities at $z\sim6$. 

\begin{figure}
    \centering
    \includegraphics[width=0.48\textwidth]{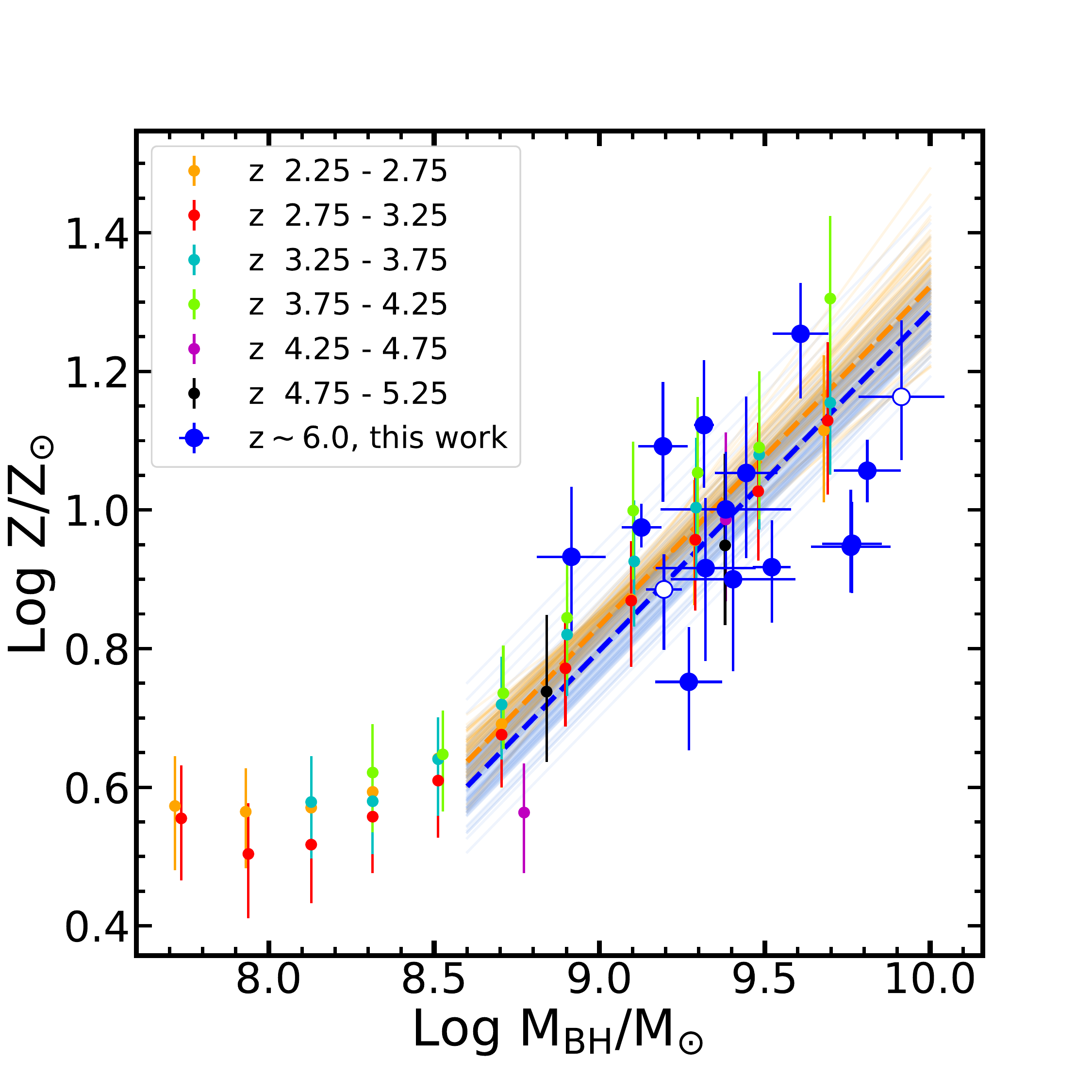}
    \caption{The BH mass-BLR metallicity relation at $z\sim6.0$ (blue) using our sample, compared with the results at $2.5<z<5.0$ \citep{Xu18}. The blue open circles represent the BH mass measured from \CIV. The thin orange lines are randomly selected linear regression fits to all points in the range of ${\rm Log}M_{\rm BH}/M_{\odot}>8.6$ in the case of $2<z<5$ \citep{Xu18}. The orange dashed line refers to the median of the distribution. The  blue thin and dashed lines are the result of this work. }
    \label{fig:mass-metal}
\end{figure}

In order to test if the above selection bias can explain the non-evolution in BLR metallicities, we compare our results with previous measurements in quasars with similar BH masses. \citet{Xu18} adopted the average metallicity indicated by \SiIV/\CIV\ and \NV/\CIV\ as their final metallicity for the BH mass-metallicity relation. They also linearly interpolate (or extrapolate) model predictions as what we did. The only difference is that they compiled the three models in \citet{Hamann02} to estimate $Z_{\rm N\,V/C\,IV}$, The models share the same assumptions of relative metal abundance scale as \citet{Nagao06}, and take the same solar abundance values from \citet{Grevesse93}. There are 16 quasars in our sample that have both \SiIV$/$\CIV\ and \NV$/$\CIV\ measurements. They are directly compared with \citet{Xu18} following the same recipes of metallicity estimation. The results are shown in Figure \ref{fig:mass-metal}. At fixed BH masses, the metallicities at $z\sim6$ from our sample are consistent with those at $2<z<5$ from \citet{Xu18}. We present a linear regression fit to both $2<z<5$ \citep{Xu18} and $z\sim6$. For the case of $2<z<5$, we choose the mass range of ${\rm Log}M_{\rm BH}/M_{\odot}>8.6$, where the mass-metallicity relation seems to be linear. The relation is expressed in the form of ${\rm Log}Z/Z_{\odot} = m\times({\rm Log}(M_{\rm BH}/M_{\odot})-{\rm Log}(M_0/M_{\odot}))+b$. $M_0$ is taken as 9.2 to make the mass range symmetric in log space to better constrain the difference in the intercepts. The best-fit slope and intercept are $0.49\pm0.05$ and $0.93\pm0.02$, respectively. The best-fit slope is different from what was presented in \citet{Xu18}, since we exclude the points below $10^{8.6}M_{\odot}$. For the case of $z\sim6$, we fix the slope to be the same in the case of $2<z<5$ because of the limited mass range and small number of points. The best-fit intercept is $0.90\pm0.04$, consistent with that of the low-redshift case within a 1$\sigma$ uncertainty. This indicates that there is no obvious evolution in the BH mass-BLR metallicity relation over $2<z<6$. Changing the mass cut range in the $2<z<5$ case or the value of $M_0$ does not change our results. Therefore, we conclude that selection bias is unable to explain the non-evolution of the quasar BLR metallicities. In fact, our comparisons with low-redshift quasar samples matched in luminosity (hence approximately in BH mass) already ruled out the selection bias as the main reason for the lack of evolution in BLR metallicities.

The BH mass-BLR metallicity relation is very different from the stellar mass-metallicity relation in star-forming galaxies. The BLR metallicities are typically $0.3\sim 1$ dex higher than those in star-forming galaxies \citep{Xu18}. In addition, there is a strong redshift evolution in the galaxy mass-metallicity relation. 
One explanation is that quasar host galaxies are distinct from star-forming galaxies in terms of metallicity and its redshift evolution \citep{Xu18}. However, studies of AGN narrow line region (NLR) metallicity do not support this explanation. The scale of the NLR is comparable to that of the host galaxy \citep[e.g.,][]{Bennert06b}. Recent studies of Type II AGNs have reached a consensus that the galaxy mass-NLR metallicity relation is similar to that for star-forming galaxies \citep{Matsouka18, Dors19}. There is also tentative evidence (see \citet{Maiolino19} and reference therein) that the AGN NLR metallicity has a similar redshift evolution as star-forming galaxies over $0<z<2$ \citep{Coil15, Dors19} (see Figure 3 in \citet{Dors19}).

An alternative explanation is that the timescale for the BLR gas to get metal enriched is very short. Although the origin of the BLR and its high metallicity is unclear, this is a viable scenario, since the gas density of the BLR is high, and the BLR can experience a rapid star formation and metal enrichment. Some studies explored in-situ star formation models in the BLR \citep{Sholsman1989, Collin99,Wang11}, where self-gravity causes instability and fragmentation of the accretion disk at outer region (e.g. beyond the self-gravity radius, $\sim 0.1$pc \citep{Collin99}). One possible picture is that these fragments will finally collapse and result in local star formation \citep{Collin99}. The supernovae explosion from these in-situ stars can inject adequate amounts of metals into the BLR and enrich the gas to as high as $10\sim20$ ${\rm Z}_{\odot}$. On the other hand, the gas mass in the BLR is only up to a few time of $10^{4}$ ${\rm M}_{\odot}$ \citep{Baldwin03}, therefore the enrichment timescale can be as short as $10^4$ yrs \citep{Juarez09}.  

Using different methods to estimate quasar metallicity other than broad emission line ratios would also be valuable. One possible approach is to use quasar intrinsic absorption lines \citep[see references in][]{Hamann99}. Many studies confirm that the metallicities indicated from either broad or narrow absorption lines, are also super solar \citep[e.g.,][]{Arav99, Arav20, Gabel06}. In our sample, many quasars exhibit diverse intrinsic absorption features. However, the typical spectral resolution and the S/N of our sample make it difficult to deploy this method.

\section{Summary}

We have performed detailed spectral analysis of 33 quasars at $5.7<z<6.4$ to study their BLR metallicities. These quasars were drawn from a sample of 50 quasars observed using Gemini GNIRS. The NIR spectra cover $0.9-2.5\mu m$. The \CIV\ or \MgII\ line flux can be robustly measured in the 33 quasars. We used different line flux ratios as metallicity diagnostics, including \SiIV$/$\CIV, \CIII$/$\CIV, \AlIII$/$\CIV, \HeII$/$\CIV, \NV$/$\CIV, and \NV$/$\HeII. The \FeII$/$\MgII\ ratios were also measured. Our main conclusions are as below.

\begin{enumerate}
    \item We compared different metallicity diagnostic line flux ratios with earlier samples measured at various redshifts. The median ratios of our sample are consistent with the luminosity-matched sample at $2<z<4.5$, suggesting no obvious redshift evolution in BLR metallicity up to $z\sim6$. The high S/N median composite spectrum from our sample confirms this non-evolution.
    
    \item We converted the observed line flux ratios to metallicities using photoionization model predictions. The typical metallicity of our sample depends on the indicator used, but is at least a few times the solar value. Our results imply the gas in the BLR is already highly enriched at $z\sim6$.
    \item We compared the \FeII$/$\MgII\ ratios with those measured for quasars at other redshifts. There is no evidence of redshift evolution in the \FeII$/$\MgII\ ratio in quasars up to $z\sim 6$, suggesting rapid star formation happened at earlier epochs.
    
    \item We found a consistent relation between the black hole mass and the BLR metallicity at $z\sim 6$ as seen in low-redshift quasars, and ruled out selection biases as the main cause for the non-evolution of quasar BLR metallicity. 
\end{enumerate}

Our results confirmed similar earlier studies with smaller sample sizes and/or less coverage of various UV broad emission lines. The confirmation of no evolution in the BLR metallicity and the Fe/$\alpha$ ratio up to $z\sim6$ strongly constrains star formation and metal enrichment in the vicinity of the SMBH. Better metallicity diagnositics can further solidify these results, and refine the metallicity measurements in individual quasars.

%%%%%%%%%%%%%%%%%%%%%%%%%%%%%%%%%%%%%%%%%%%%%%%%%%%%%%%
%%% Acknowledgements.
\begin{acknowledgements}
We are grateful to the anonymous referee for the useful comments. We thank Hiroaki Sameshima for providing their measurements, Ping Chen, Fei Xu and Hengxiao Guo for useful discussion. SW, LJ, LH, JW, ZS, and XW acknowledge support from the National Key R\&D Program of China (2016YFA0400703, 2016YFA0400702) and the National Science Foundation of China (11721303, 11890693, 11991052). YS acknowledges support from NSF grants AST-1715579 and AST-2009947. M.V. gratefully acknowledges financial support from the Independent Research Fund Denmark via grant number DFF 8021-00130. F. W acknowledges the support by NASA through the NASA Hubble Fellowship grant \#HST-HF2-51448.001-A and $\#$HF2-51434 awarded by the Space Telescope Science Institute, which is operated by the Association of Universities for Research in Astronomy, Incorporated, under NASA contract NAS5-26555.
\end{acknowledgements}

%%%%%%% Begin reference
\bibliography{ref}

%%%%%%% Begin Appendix
\begin{appendix}

\renewcommand{\thesection}{Appendix}
\section{Effects of the \FeII\ template on the \FeII$/$\MgII\ ratio} \label{sec:appendix_1}

\begin{figure}[h]
    \centering
    \includegraphics[width=0.6\textwidth]{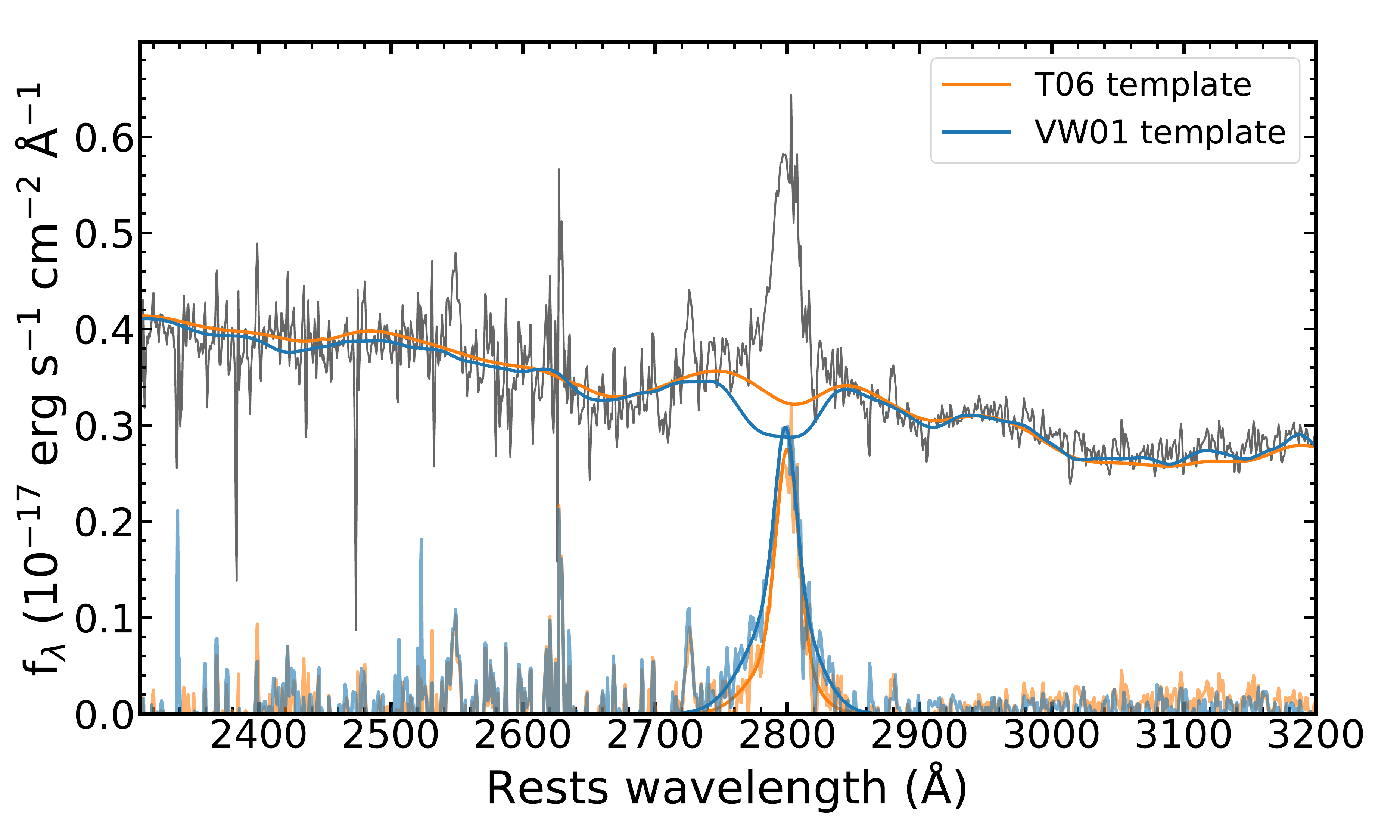}
    \caption{An example (J1250+3130) of our fits using two \FeII\ templates, VW01 (blue) and T06 (orange). The upper spectrum in black is the original spectrum, and the lower spectrum (in blue and orange) represents the pseudo-continuum (VW01 and T06, respectively) subtracted \MgII\ profile. We overplot the best-fit Gaussian models using thick blue and orange lines. The fitting results using the VW01 template have systematically lower \MgII\ fluxes than using T06.}
    \label{fig:fe2template_comparison}
\end{figure}

We investigate the \FeII$/$\MgII\ ratios using the T06 template to model \FeII. We follow the same fitting approach but replace the VW01 template with the T06 template. Figure~\ref{fig:fe2template_comparison} compares the two \FeII\ templates with an example to illustrate the differences between the fitting results. Using the T06 template produces higher \FeII\ and lower \MgII\ fluxes, thus higher \FeII$/$\MgII\ ratio. Table \ref{tab:fe2tomg2_ratios} includes \FeII$/$\MgII\ ratios measured for our sample using the T06 template.

\begin{figure}[h]
    \centering
    \includegraphics[width=0.75\textwidth]{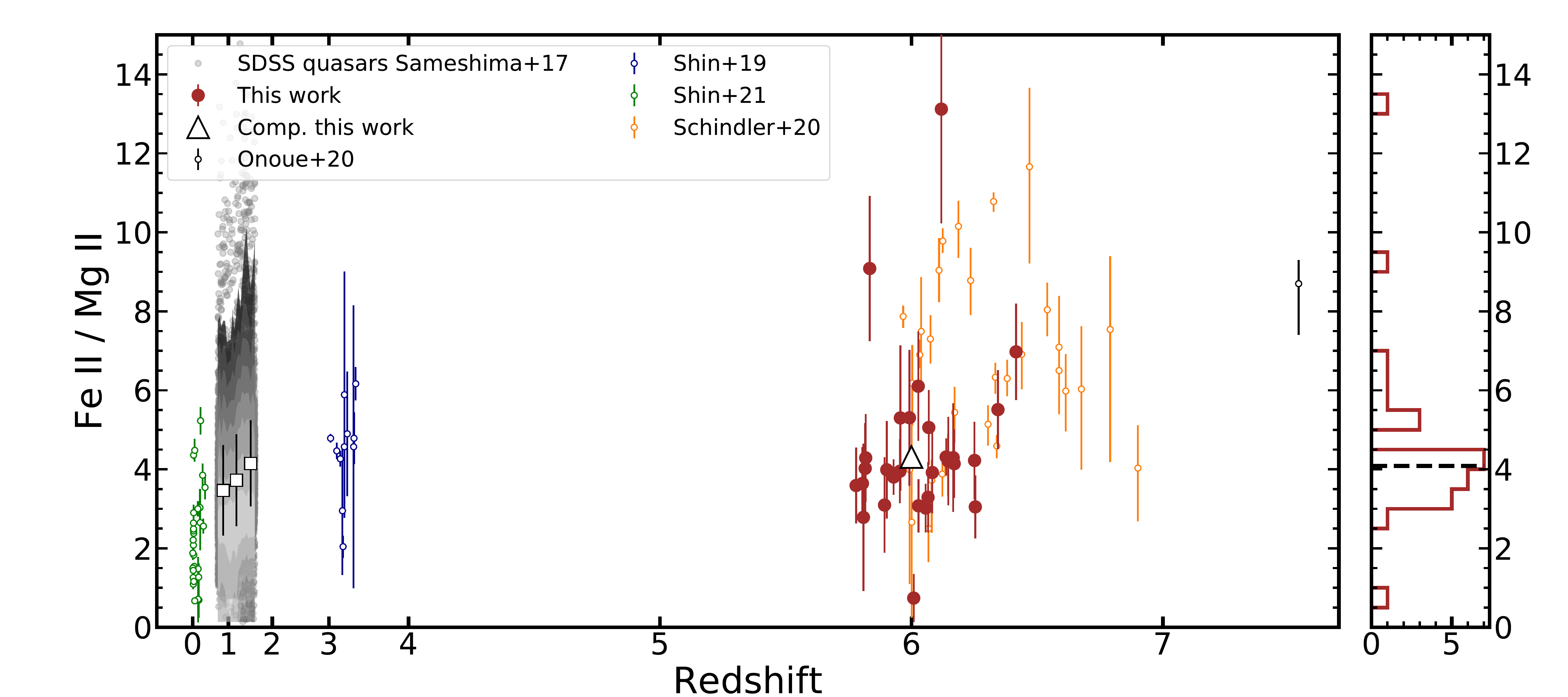}
    \caption{Same as Figure \ref{fig:fe2mg2_vs_z} but for the results derived using the T06 template. The results for SDSS quasars measured using the T06 templates are from \citet{Sameshima17}.}
    \label{fig:fe2mg2_vs_z_t06}
\end{figure}

Figure \ref{fig:fe2mg2_vs_z_t06} compares our measurements with other samples using the T06 template \citep{Sameshima17,Shin19,Shin21,Schindler20}. The median values of SDSS quasars in three redshift bins ($0.7<z<1.1$, $1.1<z<1.4$, $1.4<z<1.7$) are $3.46\pm1.15$, $3.72\pm1.16$, and $4.14\pm1.09$, respectively. Our median value, $4.08^{+1.36}_{-1.00}$, is consistent with those for low redshift SDSS quasars. In addition, the value measured from our high S/N composite spectrum (4.31$\pm0.11$) is generally in line with this result. Similar to our fiducial measurements using the VW01 \FeII\ template, we find no obvious redshift evolution in the \FeII$/$\MgII\ ratio using the T06 template. For consistency check, we compare our \FeII$/$\MgII\ measurements with other works using T06 \citep{Schindler20} for those overlapped objects.  There are three overlapped objects with \citet{Schindler20}: J0842$+$1218, J1148$+$0702, and J2310$+$1855. For two of them, the measurements are consistent within 1$\sigma$ uncertainty, while for the other object it shows large difference, indicating possible systematics in spectral decomposition or data reduction.  A uniform pipeline is needed in the future to reduce any systematics.

\end{appendix}

\end{document}